\documentclass[aps,prd,preprint,nofootinbib]{revtex4}

\usepackage{graphicx}
\usepackage{psfrag}
\usepackage{epsfig}

\begin{document}

\title{Kaon Electromagnetic Form Factor within the $k_T$ Factorization
Formalism and It's Light-Cone Wave Function}
\author{Xing-Gang Wu$^{1}$ \footnote{email: wuxg@cqu.edu.cn} and
Tao Huang$^{2}$\footnote{email: huangtao@mail.ihep.ac.cn}}
\address{$^1$Department of Physics, Chongqing University, Chongqing 400044,
P.R. China\\ $^2$Institute of High Energy Physics, Chinese Academy
of Sciences, P.O.Box 918(4), Beijing 100049, P.R. China}

\begin{abstract}
We present a systematical study on the kaon electromagnetic form
factors $F_{K^{\pm},K^0,\bar{K}^0}(Q^2)$ within the $k_T$
factorization formalism, where the transverse momentum effects, the
contributions from the different helicity components and different
twist structures of the kaon light-cone (LC) wave function are
carefully analyzed for giving a well understanding of the hard
contributions at the energy region where pQCD is applicable. The
right power behavior of the hard contribution from the higher
helicity components and from the higher twist structures can be
obtained by keeping the $k_T$ dependence in the hard amplitude. Our
results show that the $k_T$ dependence in LC wave function affects
the hard and soft contributions substantially and the
power-suppressed terms (twist-3 and higher helicity components) make
an important contribution below $Q^2\sim several\; GeV^2$ although
they drop fast as $Q^2$ increasing. The parameters of the proposed
model wave function can be fixed by the first two moments of its
distribution amplitude and other conditions. By varying the first
two moments $a^K_1(1GeV)$ and $a^K_2(1GeV)$ with the region of
$0.05\pm0.02$ and $0.10\pm 0.05$ respectively, we find that the
uncertainty of the kaon electromagnetic form factor is rather small. \\


\end{abstract}
\maketitle

\section{Introduction}

The electromagnetic form factors provide useful information
concerning the internal structures of the mesons, and they also
provide useful platforms to check the rightness of the perturbative
QCD (pQCD) theory. Recently, the electromagnetic form factor
$F_\pi(Q^2)$ has been restudied in Refs.\cite{piem1,piem2,piem3}. It
was shown that when all the power suppressed contributions, which
include higher order in $\alpha_s$, higher helicities and higher
twists in the light-cone (LC) wave function, and etc., have been
taken into account, then the hard contributions can fit the present
experimental data well at the energy region where pQCD is
applicable. By comparison the behavior of the kaon electromagnetic
form factor is less certain both experimentally and theoretically.
The kaon is composed by two quarks with different quark masses,
therefore it becomes a little more complicated to obtain its LC wave
functions and to compute its electromagnetic form factor. For
example, the kaon electromagnetic form factor has been studied in
Refs.\cite{softkaon,maboqiang} in the light-cone quark model only
within the soft region. Here we shall present a systematical study
on the charged/neutral kaon electromagnetic form factors in the
intermediate and large energy region within the $k_T$ factorization
formalism by properly taking the $SU_f(3)$ breaking effects into
account.

The kaon electromagnetic form factor can be obtained through the
definition
\begin{equation}
\langle K(p^{\prime})|J_{\mu}|K(p)\rangle=(p+p^{\prime})_{\mu}
F_{K}(Q^2),
\end{equation}
where $K$ stands for $K^{\pm}$, $K^0$ and $\bar{K}^0$ respectively,
the vector current $J_{\mu}=\sum_i e_i\bar{q}_i\gamma_{\mu}q_{i}$,
with the quark flavor $i$ and the relevant electric charge $e_i$.
The momentum transfer $q^2=-Q^2=(p-p^{\prime})^2$ is restricted in
the space-like region. In the LC quantization and by using the
Drell-Yan-West ($q^+=0$) frame \cite{dy}, the kaon electromagnetic
form factor can generally be expressed as
\begin{equation}\label{basic:form}
F_{K} (Q^2) = \hat\Psi \otimes \hat\Psi
=\sum_{n,\lambda_i}\sum_j{e_j}\int [dx_i] [d\mathbf{k}_{i\perp}]_n
\Psi^*_n(x_i,\mathbf{k}_{i\perp},\lambda_i)
\Psi_n(x_i,\mathbf{k}_{i\perp}+\delta_iq_\perp,\lambda_i),
\label{fpi0}
\end{equation}
where the summation extends over all quark/gluon Fock states which
have a non-vanishing overlap with the kaon, $e_j$ is the electric
charge of the struck quark, $\Psi_n$ are the corresponding wave
functions which describe both the low and the high momentum partons,
$[dx_i][d\mathbf{k}_{i\perp}]_n$ is the relativistic measure within
the $n$-particle sector and $\delta_i=(1-x_i)$ or $(-x_i)$ depending
on whether $i$ refers to the struck quark or a spectator,
respectively.

Similar to the pionic case, it can be found that the nominal power
law contribution to $F_{K}(Q^2)$ as $Q\rightarrow \infty$ is
$F_{K}(Q^2)\sim 1/(Q^2)^{n-1}$ in the light-cone gauge $(A^+=0)$
\cite{brodsky}, under the condition that $n$ quark or gluon
constituents are forced to change direction. Thus only the
$q\bar{q}$ component of $\Psi^{((1-x)Q)}(x,{\bf k}_\perp,\lambda)$
contributes at the leading $1/Q^2$. For the large $Q^2$ region, the
hard contribution to the kaon electromagnetic form factor can be
written as
\begin{eqnarray}
&&F_{K}(Q^2)=\nonumber\\
&&\sum_j{e_j}\int [dx][dy][d^2{\bf k_{\perp}}][d^2{\bf l_{\perp}}]
\Psi^{*(1-x)Q}(x,{\bf k_{\perp}},\lambda) T_{H}(x,y,{\bf
q_{\perp},{\bf k_{\perp}},{\bf
l_{\perp}},\lambda,\lambda^{\prime}})\Psi^{(1-y)Q}(y,{\bf
l_{\perp}},\lambda^{\prime})+ \cdots,\label{factor}
\end{eqnarray}
where the ellipses represent the higher Fock states' contributions,
$[dx]=dx_1d_2\delta(1-x_1-x_2)$ and $[d^2{\bf k_{\perp}}]=d^2{\bf
k_{\perp}}/16\pi^3$. $\Psi^{((1-x)Q)}(x,{\bf k}_\perp,\lambda)$ is
the valence Fock-state LC wave function with helicity $\lambda$ and
with a cut-off on $|{\bf k}_\perp|$ that is of order $(1-x)Q$. Such
a cut-off on $|{\bf k}_\perp|$ is necessary to ensure that the wave
function is only responsible for the lower momentum region. And the
hard scattering amplitude $T_H$ contains all two-particle
irreducible amplitudes for $\gamma^*+q\bar{q}\to q\bar{q}$.

The LC wave function provides useful links between the hadronic
phenomena in QCD at large distance (non-perturbative) and small
distance (perturbative). A LC wave function is a localized
stationary solution of the LC schr\"{o}dinger equation $i\partial
|\Psi(\tau)\rangle =H_{LC}|\Psi(\tau)\rangle$ \cite{lbhm,lb}, which
describes the evolution of a state $|\Psi(\tau)\rangle$ on the LC
time $\tau\equiv x^{+}=x^{0}+x^{3}$ in the physical LC gauge
$A^{+}=0$. For the valence quark state of the kaon, its LC wave
functions can be defined in terms of the bilocal operator matrix
element \cite{BenekeFeldmann},
\begin{eqnarray}
&&\langle K(p)|\bar q_\beta(z) q_\alpha|0\rangle=\nonumber\\
&&\frac{i \sqrt{6}}{2} \Bigg\{ \slash\!\!\!p\,\gamma_5\,\Psi_{K}(x,
\mathbf{k_\perp})- \mu_K\gamma_5 \left[\Psi_p(x, \mathbf{k_\perp})
-i\sigma_{\mu\nu}\left(n^{\mu}\bar{n}^{\nu}\,\frac{\Psi_{\sigma}'(x,
\mathbf{k_\perp})}{6}-p^\mu\,\frac{\Psi_\sigma(x,
\mathbf{k_\perp})}{6}\, \frac{\partial}{\partial
\mathbf{k}_{\perp\nu}} \right)\right]
\Bigg\}_{\alpha\beta},\label{waveoperator}
\end{eqnarray}
where $\mu_K$ is the phenomenological parameter:
$\mu_K=M_K^2/(m_s+m_u)$ for $K^\pm$ and $\mu_K=M_K^2/(m_s+m_d)$ for
$K^0$ or $\bar{K}^0$ respectively, which is a scale characterized by
the chiral perturbation theory. $\Psi_{K}(x,\mathbf{k_{\perp}})$ is
the leading twist (twist-2) wave function,
$\Psi_p(x,\mathbf{k_{\perp}})$ and
$\Psi_{\sigma}(x,\mathbf{k_{\perp}})$ are sub-leading twist
(twist-3) wave functions that correspond to the pseudo-scalar
structure and the pseudo-tensor structure respectively. The wave
function $\Psi(x,\mathbf{k}_\perp)$ ($\Psi$ stands for $\Psi_K$,
$\Psi_p$ and $\Psi_\sigma$ respectively) satisfies the normalization
condition
\begin{equation}\label{norm}
\int^1_0 dx \int\frac{d^{2}{\bf
k}_{\perp}}{16\pi^3}\Psi(x,\mathbf{k}_\perp)=\frac{f_K}{2\sqrt{6}},
\end{equation}
where the decay constant $f_K=160MeV$ \cite{pdg}. The distribution
amplitude (DA) $\phi(x)$ and the wave function
$\Psi(x,\mathbf{k_{\perp}})$ are related by
\begin{equation}\label{phipsi}
\phi(x)=\frac{2\sqrt{6}}{f_K}\int_{|\mathbf{k}_\perp|<\mu_f}
\frac{d^2\mathbf{k_\perp}}{16\pi^3}\Psi(x,\mathbf{k_\perp}).
\end{equation}

Non-leading perturbative contributions to the kaon electromagnetic
form factor include the higher order in $\alpha_s$, higher
helicities and higher twists in the LC wave function, and etc.
Similar to the pionic case \cite{piem1,piem2}, it is substantial to
take $k_T$ dependence in the wave function into account and to keep
the transverse momentum dependence fully in the hard scattering
amplitude in the $k_T$ factorization formalism within the LC
framework. In present paper, we shall calculate all the helicity
components' contributions to the kaon electromagnetic form factor
within the LC pQCD framework, which is consistent with the using of
LC wave function. Another important power correction is from the
higher twist structures in the kaon DA. The end-point singularity
becomes more serious for the higher twist structures, because the
asymptotic behavior of the twist-3 DAs, especially
$\phi^{as}_p(x)=1$, so the calculations for these higher twist
contributions have more uncertainty than that for the leading twist.
It means that one should use the twist-3 wave function with a better
behavior in the end-point region than that of the asymptotic one so
as to give a more reliable estimation of the higher twist
structures' contribution. The Brodsky-Huang-Lepage (BHL)
prescription provides a useful way to construct a wave function with
better end-point behavior \cite{bhl}, we shall adopt it to construct
the kaon LC wave functions for the present purpose, and then we
discuss its uncertainty for the kaon electromagnetic form factor.
The $SU_f(3)$-breaking effects shall also be included for
constructing the kaon LC wave function.

The reminder of the paper is organized as follows. Sec.II is devoted
to present the main properties of the kaon electromagnetic form
factor and the formulae for the twist-2 and twist-3 contributions to
the kaon electromagnetic form factor within the $k_T$ factorization
approach. Numerical results for the kaon electromagnetic form factor
are presented in Sec.III. The last section is reserved for
conclusion and summary.

\section{Kaon electromagnetic form factor within the $k_T$ factorization
formalism}

Because, $K^+=u\bar{s}$ and $K^-=s\bar{u}$, one can find that
$F_{K^+}=-F_{K^-}$ according to Eq.(\ref{basic:form}). Similarly,
since $K^0=d\bar{s}$ and $\bar{K}^0=s\bar{d}$, it can be found that
$F_{K^0}=-F_{\bar{K}^0}$. So, we only need to calculate the $K^+$
and $K^0$ form factors, where $e_1=2/3$, $e_2=1/3$, $m_1=m_u$ and
$m_2=m_s$ for $F_{K^+}$ and $e_1=-1/3$, $e_2=1/3$, $m_1=m_d$ and
$m_2=m_s$ for $F_{K^0}$ respectively. Here $m_u=m_d\neq m_s$ stand
for the light constitute quark masses. Further more, in doing the
calculation of the hard scattering amplitude with the $k_T$
factorization formulism, we shall treat the current quark masses of
$u$, $d$ and $s$ to be zero due to their smallness in comparison to
the involved hard scale. Then the calculation procedure for the hard
scattering amplitude is the same as that of the pionic case
\cite{piem2,piem1}.

\subsection{Formulae for the twist-2 contribution to the kaon electromagnetic form
factor}

\begin{figure}
\includegraphics[width=0.7\textwidth]{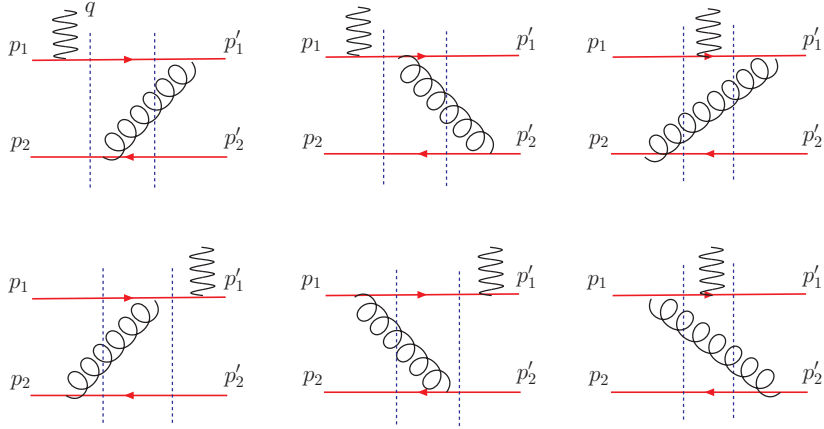}
\caption{Six leading order time-ordered Feynman diagrams for the
hard scattering amplitude $T_H$, where
$p_1=(x_1,\mathbf{k}_{\perp})$, $p_2=(x_2,-\mathbf{k}_{\perp})$,
$p_1^{\prime}=(y_1,y_1\mathbf{q}_{\perp}+\mathbf{l}_{\perp})$,
$p_2^{\prime}=(y_2,y_2\mathbf{q}_{\perp}-\mathbf{l}_{\perp})$. }
\label{feyn}
\end{figure}

\begin{table}
\caption{Full form of the LC wave function
$\Psi(x,\mathbf{k_\perp,\lambda})=\varphi(x,\mathbf{k_\perp})\chi$.
$\Psi(x,\mathbf{k_\perp},\lambda)$ stands for
$\Psi_K(x,\mathbf{k_\perp},\lambda)$,
$\Psi_p(x,\mathbf{k_\perp},\lambda)$ and
$\Psi_\sigma(x,\mathbf{k_\perp},\lambda)$, respectively.}
\begin{center}
\begin{tabular}{|c||c|c|}
\hline\hline ~~~$\lambda_1\lambda_2$~~~ & ~~~$\uparrow\uparrow$~~~&
~~~$\uparrow\downarrow$~~~ \\
\hline $\Psi_{\lambda_1\lambda_2}(x,\mathbf{k_\perp},\lambda)$ &
$-\frac{(a_1+a_2)(k_x-i k_y)} {[2(a^2_1+\mathbf{k}^2_\perp)
(a^2_2+\mathbf{k}^2_\perp)]^{1/2}}\varphi(x,\mathbf{k_\perp})$ &
$\frac{(a_1 a_2-\mathbf{k}_\perp^2)} {[2(a^2_1+\mathbf{k}^2_\perp)
(a^2_2+\mathbf{k}^2_\perp)]^{1/2}}\varphi(x,\mathbf{k_\perp})$ \\
\hline\hline
~~~$\lambda_1\lambda_2$~~~ &~~~$\downarrow\uparrow$~~~ & ~~~$\downarrow\downarrow$~~~ \\
\hline $\Psi_{\lambda_1\lambda_2}(x,\mathbf{k_\perp},\lambda)$ &
$-\frac{(a_1 a_2-\mathbf{k}_\perp^2)} {[2(a^2_1+\mathbf{k}^2_\perp)
(a^2_2+\mathbf{k}^2_\perp)]^{1/2}}\varphi(x,\mathbf{k_\perp})$ &
$-\frac{(a_1+a_2)(k_x+i k_y)} {[2(a^2_1+\mathbf{k}^2_\perp)
(a^2_2+\mathbf{k}^2_\perp)]^{1/2}}\varphi(x,\mathbf{k_\perp})$ \\
\hline\hline
\end{tabular}
\label{tab1}
\end{center}
\end{table}

In the intermediate and large energy region, one can apply the pQCD
approach and use the valence Fock state to estimate the kaon
electromagnetic form factor since the applicability of pQCD in the
intermediate and large energy region has been proved by involving
the $k_T$ dependence \cite{lsh}. The lowest-order contribution for
the hard scattering amplitude $T_H$ comes from the one-gluon
exchange Feynman diagrams as shown in Fig.(\ref{feyn}). To
simplicity our notations, we separate the spin-space wave function
$\chi^K(x,{\bf k}_{\perp},\lambda)$ out from the whole LC wave
function, i.e., $\Psi^{(1-x)Q}(x,{\bf
k_{\perp}},\lambda)\to\chi^K(x,{\bf
k}_{\perp},\lambda)\varphi^{(1-x)Q}(x,{\bf k_{\perp}})$, where
$\Psi^{((1-x)Q)}(x,{\bf k}_\perp)$ is the light-cone wave function
of the valence Fock state with a cut-off $|{\bf k}_\perp|$ of order
$(1-x)Q$ and the spin space wave function $\chi^K(x,{\bf
k}_{\perp},\lambda)$ that comes from the spin space Wigner rotation
can be found in Ref.\cite{hms} \footnote{Setting $m_1=m_2=m_q$, we
return to the results of the case of pion.}, which is given in
TAB.\ref{tab1}. One can combine the spin-space wave function
$\chi^K(x,{\bf k}_{\perp},\lambda)$ into the original $T_H$ to form
a new one, i.e.,
\begin{eqnarray}
T_H &=& (e_1\xi_1 +e_2\xi_2)
T_H^{(\lambda_1+\lambda_2=0)}(\uparrow\downarrow
\rightarrow\uparrow\downarrow)+ (e_1\xi_1 +e_2\xi_2)
T_H^{(\lambda_1+\lambda_2=0)}(\downarrow\uparrow
\rightarrow\downarrow\uparrow)+\nonumber\\
& &  (e_1\xi'_1 +e_2\xi'_2) T_H^{(\lambda_1+\lambda_2=
1)}(\uparrow\uparrow \rightarrow\uparrow\uparrow)+(e_1\xi^{\prime
*}_1 +e_2\xi^{\prime *}_2) T_H^{(\lambda_1+\lambda_2=-
1)}(\downarrow\downarrow \rightarrow\downarrow\downarrow)\ ,
\end{eqnarray}
where $\lambda_{1,2}$ are the helicities for the (initial or final)
kaon's two constitute quarks respectively. It is found that there is
no hard scattering amplitude with quark and antiquark helicities
being changed due to the fact that the quark helicity is conserved
at each quark-gluon (photon)-quark vertex in the limit of vanishing
quark mass. $e_i$ is the electric charge of the struck quark,
$\xi_i$ and $\xi'_i$ are coefficients derived from $\chi^K(x,{\bf
k}_{\perp},\lambda)$,
\begin{eqnarray}
\xi_1 &=& \frac{(a_1 a_2-\mathbf{k}_\perp^2)(a'_1
a'_2-\mathbf{l}_\perp^{2})} {2[(a^2_1+\mathbf{k}^2_\perp)
(a^2_2+\mathbf{k}^2_\perp)(a^{\prime 2}_1+
\mathbf{l}^2_\perp)(a^{\prime 2}_2+\mathbf{l}^2_\perp)]^{1/2}}, \nonumber\\
\xi'_1 &=& \frac{(a_1+a_2)(a'_1 + a'_2)(\mathbf{k}_\perp \cdot
\mathbf{l}_\perp+i\mathbf{k}_\perp \times \mathbf{l}_\perp )}
{2[(a^2_1+\mathbf{k}^2_\perp) (a^2_2+\mathbf{k}^2_\perp)(a^{\prime
2}_1+ \mathbf{l}^2_\perp)(a^{\prime 2}_2+\mathbf{l}^2_\perp)]^{1/2}}, \nonumber\\
\xi_2 &=& \frac{(b_1 b_2-\mathbf{k}_\perp^2)(b'_1
b'_2-\mathbf{l}_\perp^{2})} {2[(b^2_1+\mathbf{k}^2_\perp)
(b^2_2+\mathbf{k}^2_\perp)(b^{\prime 2}_1+
\mathbf{l}^2_\perp)(b^{\prime 2}_2+\mathbf{l}^2_\perp)]^{1/2}}
\end{eqnarray}
and
\begin{eqnarray}
\xi'_2 &=& \frac{(b_1+b_2)(b'_1 + b'_2)(\mathbf{k}_\perp \cdot
\mathbf{l}_\perp+i\mathbf{k}_\perp \times \mathbf{l}_\perp )}
{2[(b^2_1+\mathbf{k}^2_\perp) (b^2_2+\mathbf{k}^2_\perp)(b^{\prime
2}_1+ \mathbf{l}^2_\perp)(b^{\prime 2}_2+
\mathbf{l}^2_\perp)]^{1/2}} \ ,
\end{eqnarray}
where
\begin{eqnarray*}
&& a_1=x M_a+m_1, a_2=(1-x)M_a+m_2 \\
&& a'_1=y M'_a+m_1, a'_2=(1-y)M'_a+m_2 \\
&& b_1=x M_b+m_2, b_2=(1-x)M_b+m_1 \\
&& b'_1=y M'_b+m_2, b'_2=(1-y)M'_b+m_1  \\
&& M_a^2=\frac{m_1^2+\mathbf{k}_\perp^2}{x}
+\frac{m_2^2+\mathbf{k}_\perp^2}{1-x}, M_a^{\prime
2}=\frac{m_1^2+\mathbf{l}_\perp^2}{y}
+\frac{m_2^2+\mathbf{l}_\perp^2}{1-y} \\
&& M_b^2=\frac{m_2^2+\mathbf{k}_\perp^2}{x}
+\frac{m_1^2+\mathbf{k}_\perp^2}{1-x}, M_b^{\prime
2}=\frac{m_2^2+\mathbf{l}_\perp^2}{y}
+\frac{m_1^2+\mathbf{l}_\perp^2}{1-y}
\end{eqnarray*}
Consequently, the above coefficients can be further simplified as
\begin{eqnarray}
\xi_1 &=& \frac{[m_{1}(1-x)+m_{2}x][m_{1}(1-y)+m_{2}y]}
{2\sqrt{k^2_\perp+\left[m_{1}(1-x)+m_{2}x\right]^2}
\sqrt{l^2_\perp+\left[m_{1}(1-y)+m_{2}y\right]^2}}, \nonumber\\
\xi'_1 &=& \frac{(\mathbf{k}_\perp \cdot
\mathbf{l}_\perp+i\mathbf{k}_\perp \times \mathbf{l}_\perp )}
{2\sqrt{k^2_\perp+\left[m_{1}(1-x)+m_{2}x\right]^2}
\sqrt{l^2_\perp+\left[m_{1}(1-y)+m_{2}y\right]^2}}, \nonumber\\
\xi_2 &=& \frac{[m_{2}(1-x)+m_{1}x][m_{2}(1-y)+m_{1}y]}
{2\sqrt{k^2_\perp+\left[m_{2}(1-x)+m_{1}x\right]^2}
\sqrt{l^2_\perp+\left[m_{2}(1-y)+m_{1}y\right]^2}}
\end{eqnarray}
and
\begin{eqnarray}
\xi'_2 &=& \frac{(\mathbf{k}_\perp \cdot
\mathbf{l}_\perp+i\mathbf{k}_\perp \times \mathbf{l}_\perp )}
{2\sqrt{k^2_\perp+\left[m_{2}(1-x)+m_{1}x\right]^2}
\sqrt{l^2_\perp+\left[m_{2}(1-y)+m_{1}y\right]^2}} \ .
\end{eqnarray}
Schematically, the total hard scattering amplitude can be written as
\begin{displaymath}
T_H=(e_1\xi_1 +e_2\xi_2)T_{H}^{(\lambda_1+\lambda_2=0)}+
[e_1(\xi_1+\xi_1^{\prime *})+e_2(\xi_2+\xi_2^{\prime
*})]T_{H}^{(\lambda_1+\lambda_2=\pm 1)}
\end{displaymath}
with
\begin{eqnarray}\label{hardamp1}
T_{H}^{(\lambda_1+\lambda_2=0)}&=&\frac{16\pi {C_F}
{\alpha_s(\mu_f^2)}}{ ( 1- x )( 1-y )x y} \times((( x-1
)\mathbf{q_\perp}^2 -2\mathbf{k_\perp}\cdot \mathbf{q_\perp})( 2
\mathbf{l_\perp}\cdot\mathbf{q_\perp} +
(y -1)\mathbf{q_\perp}^2 ))^{-1}\nonumber\\
& & ((x-1) ( 2 \mathbf{l_\perp} \cdot\mathbf{q_\perp} +   (y
-1)\mathbf{q_\perp}^2 )-2 (y -1)
\mathbf{k_\perp} \cdot\mathbf{q_\perp})^{-1}\times\nonumber\\
& & \Bigg[ 2 ( y-1) y ( 1 - y + x ( 2 y-1 ) )
{(\mathbf{k_\perp}\cdot\mathbf{q_\perp})}^2 + ( x-1) x (
2\mathbf{l_\perp}\cdot\mathbf{q_\perp} + ( y-1
)\mathbf{q_\perp}^2)\cdot\nonumber\\
& & ( ( 1 - y + x ( 2y-1 ) )
(\mathbf{l_\perp}\cdot\mathbf{q_\perp})+ 2 ( x-1 ) ( y-1) y
\mathbf{q_\perp}^2 ) - \nonumber\\
& & ( x-1) ( y-1) y (\mathbf{k_\perp}\cdot\mathbf{q_\perp})\cdot( 8
x (\mathbf{l_\perp}\cdot\mathbf{q_\perp}) + ( 1 - y + x ( 6y-5))
\mathbf{q_\perp}^2 ) \Bigg],
\end{eqnarray}
and
\begin{eqnarray}\label{hardamp2}
T_{H}^{(\lambda_1+\lambda_2=\pm 1)}&=& \frac{8 \pi {C_F}
{\alpha_s(\mu_f^2)}}{ ( 1- x )( 1-y )x y } \times((( x-1
)\mathbf{q_\perp}^2-2\mathbf{k_\perp}\cdot\mathbf{q_\perp}) (2
\mathbf{l_\perp}\cdot\mathbf{q_\perp} +  (y -1)
\mathbf{q_\perp}^2 ))^{-1}\nonumber\\
& & ((x-1) ( 2 \mathbf{l_\perp} \cdot\mathbf{q_\perp} +   (y
-1)\mathbf{q_\perp}^2 )-2 (y -1) \mathbf{k_\perp}
\cdot\mathbf{q_\perp})^{-1}\Bigg[ 2( x-1) x {(\mathbf{l_\perp}
\cdot\mathbf{q_\perp})}^2+ \nonumber\\
& & ( y-1) ( 2y{(\mathbf{k_\perp} \cdot\mathbf{q_\perp})}^2 + ( x-1)
(x(\mathbf{l_\perp} \cdot\mathbf{q_\perp})- y
(\mathbf{k_\perp}\cdot\mathbf{q_\perp})) \mathbf{q_\perp}^2) \Bigg],
\end{eqnarray}
where the scale $\mu_f^2=Q^2$. It can be found that the leading
contribution from the higher helicity components is of order
$1/Q^4$, which is next-to-leading contribution compared to that of
the ordinary helicity components.

With the help of Eq.(\ref{factor}), we can obtain the leading-twist
hard part contribution to the kaon form factor. And after
integrating over the azimuth angles for $\mathbf{k}_{\perp}$ and
$\mathbf{l}_{\perp}$, we obtain the contribution from the usual
helicity components $(\lambda_1+\lambda_2=0)$,
\begin{eqnarray}\label{hel0f}
F^{twist2,(\lambda_1+\lambda_2=0)}_{K}(Q^2)&=&\int dx dy d\eta_1
d\eta_2\frac{(e_1\xi_1) {C_F}
\alpha_s(\mu_f^2)|\mathbf{k}_{\perp}||\mathbf{l}_{\perp}|}{32\pi^3 x
y}\varphi(x,\mathbf{k}_\perp)\varphi^{*}(y,\mathbf{l}_\perp)\times\nonumber\\
& &\!\!\!\!\!\!\!\!\!\!\!\!\!\!\!\!\!\!\!\!\!\!
\!\!\!\!\!\!\!\!\!\!\!\!\!\!\!\!\!\!\!\!\!\! \left[ \frac{x( x + y-1
- 2xy ) } {( 1 - x ) \sqrt{1 - \eta_1^2}} +\frac{y( x + y -1- 2xy )
} {( 1 - y ) \sqrt{1 - \eta_2^2}} + \frac{x + y- x^2 - y^2} {( 1- x
) (1-y )\sqrt{1 -\eta_1^2} \sqrt{1 - \eta_2^2}}
\right]\nonumber\\
&& \!\!\!\!\!\!\!\!\!\!\!\!\!\!\!\!\!\!\!\!\!\!
\!\!\!\!\!\!\!\!\!\!\!\!\!\!\!\!\!\!\!\!\!\! +\Bigg\{
e_1\leftrightarrow e_2,\; m_1\leftrightarrow m_2 \Bigg\}
\end{eqnarray}
and the contribution from the higher helicity components
$(\lambda_1+\lambda_2=\pm 1)$,
\begin{eqnarray}
F^{twist2,(\lambda_1+\lambda_2=\pm 1)}_{K}(Q^2)&=&-\int dx dy
d\eta_1 d\eta_2 \frac{(e_1\xi_3) {C_F}\alpha_s(\mu_f^2)|
\mathbf{k}_{\perp}| |\mathbf{l}_{\perp}|}{64 \pi^3 x y
}\varphi(x,\mathbf{k}_\perp)\varphi^{*}(y,\mathbf{l}_\perp)\nonumber\\
& & \times\left[( x + y - 2xy ) \frac{\left( 1 - \sqrt{1 -
\eta_1^2}\right) \left( 1 - \sqrt{1 - \eta_2^2}\right)}
{(1-x)(1-y)\eta_1\eta_2 \sqrt{1 -\eta_1^2}\sqrt{1 -\eta_2^2}}
\right]\nonumber\\
&& +\Bigg\{ e_1\leftrightarrow e_2,\; m_1\leftrightarrow m_2
\Bigg\}\ , \label{hel1f}
\end{eqnarray}
where without loss of generality, we have implicitly assumed that
the radial kaon wave function $\varphi(x,{\bf k_{\perp}})$ depending
on $\mathbf{k}_\perp$ through $k^2_\perp$ only, i.e. $\varphi(x,{\bf
k_{\perp}})=\varphi(x,k_{\perp}^2)$. The terms in the big brace are
obtained by transforming the terms out of the brace through the
transformation $e_1\leftrightarrow e_2$ and $m_1\leftrightarrow
m_2$. $\xi_3=\frac{|\mathbf{k}_\perp| |\mathbf{l}_\perp|}
{\sqrt{k^2_\perp+\left[m_{1}(1-x) +m_{2}x\right]^2} \sqrt{l^2_\perp+
\left[m_{1}(1-y)+ m_{2}y\right]^2}}$,
$|\mathbf{k}_{\perp}|=Q(1-x)\eta_1/2$ and
$|\mathbf{l}_{\perp}|=Q(1-y)\eta_2/2$, with $\eta_{1,2}$ in the
range of $(0,\ 1)$. An overall minus sign in Eq.(\ref{hel1f})
implies that the higher helicity components shall always {\it
suppress} the contribution from the usual helicity components.

\subsection{Formulae for the twist-3 contributions to the kaon electromagnetic form
factor}

The end-point singularity becomes more serious for the higher twist
structures, because the asymptotic behavior of the twist-3 DAs,
especially $\phi^{as}_p(x)=1$, so the calculations for these higher
twist contributions have more uncertainty than that for the leading
twist. As has been pointed out in Ref.\cite{threshold1}, after
including the parton transverse momenta, large double logarithmic
corrections $\alpha_s \ln^2 k_\perp$ and $\alpha_s \ln^2x$ appear in
higher order radiative corrections and can be summed up to all
orders. The relevant Sudakov form factors from both $k_\perp$ and
the threshold resummation can cure the endpoint singularity and then
the main contribution comes from the perturbative regions. For the
present purpose, it is convenient to transform the kaon form factor
into the compact parameter $\mathbf{ b}$ space. In the large $Q^2$
region, by considering only the lowest valence quark state of the
kaon and by doing the Fourier transformation of the wave function
with the formula,
\begin{displaymath}
\Psi(x_i,{\bf k}_{\perp};\mu_f)=\int \frac{d^{2}{\bf b}}{(2\pi)^2}
e^{-i{\bf b}\cdot {\bf k_\perp}} \hat{\Psi}(x_i,{\bf b};\mu_f),
\end{displaymath}
we can transform the kaon electromagnetic form factor into the
compact parameter $\mathbf{ b}$ space,
\begin{eqnarray}
\label{twistkaon} F_{K}(Q^2)&=&\int[dx_id{\bf b}][dy_jd{\bf h}]
\hat{\Psi}(x_i,{\bf b};\mu_f)\hat{T}(x_i,{\bf b};y_j,{\bf
h};\mu_f)\hat{\Psi}(y_j,{\bf h};\mu_f)
\times S_t(x_i)S_t(y_j)\times\nonumber\\
& & \exp(-S(x_i,y_j,Q,\mathbf{b},\mathbf{h};\mu_f)),
\end{eqnarray}
where $\hat{\mu}_f=\ln(\mu_f/\Lambda_{QCD})$, $[dx_id{\bf
b}]=dx_1dx_2d^2\mathbf{b} \delta(1-x_1-x_2)/(16\pi^3)$ and the hard
kernel
\begin{displaymath}
\hat{T}(x_i,{\bf b};y_j,{\bf h};\mu_f)=\int \frac{d^{2}{\bf
k}_\perp}{(2\pi)^2}\frac{d^{2}{\bf l}_\perp}{(2\pi)^2}e^{-i{\bf
b}\cdot {\bf k_\perp}-i{\bf h}\cdot {\bf l_\perp}} T(x_i,{\bf
k}_{\perp i};y_j,{\bf l}_{\perp j};\mu_f).\nonumber
\end{displaymath}
The factor $\exp(-S(x_i,y_j,Q,\mathbf{b},\mathbf{h};\mu_f))$
contains the Sudakov logarithmic corrections and the renormalization
group evolution effects of both the wave functions and the hard
scattering amplitude,
\begin{equation}
S(x_1,y_1,Q,\mathbf{b},\mathbf{h};\mu_f)= \left[\left(\sum_{i=1}^2
s(x_i,b,Q)+\sum_{j=1}^{2}s(y_j,h,Q) \right)
-\frac{1}{\beta_{1}}\ln\frac{\hat{\mu}_f}{\hat{b}}
-\frac{1}{\beta_{1}}\ln\frac{\hat{\mu}_f}{\hat{h}} \right],
\end{equation}
where ${\hat b} \equiv  {\rm ln}(1/b\Lambda_{QCD})$, ${\hat h}
\equiv {\rm ln}(1/h\Lambda_{QCD}) $ and $s(x,b,Q)$ is the Sudakov
exponent factor, whose explicit form up to next-to-leading log
approximation can be found in Ref.\cite{liyu}. In
Eq.(\ref{twistkaon}), $S_t(x_i)$ and $S_t(y_i)$ come from the
threshold resummation effects and the exact form of each involves
one parameter integration \cite{kls}. In order to simplify the
numerical calculations, we take a simple parametrization proposed in
Ref.\cite{kls},
\begin{equation}
S_t(x)=\frac{2^{1+2c}\Gamma(3/2+c)}{\sqrt{\pi}\Gamma(1+c)}
[x(1-x)]^c\;,
\end{equation}
where the parameter $c$ is determined around $0.3$.

With the help of the above equations, we obtain the formula for the
twist-3 contributions to the kaon electromagnetic form factor,
\begin{eqnarray}\label{finalQ}
F^{twist3}_{K}(Q^2)&=&\frac{128\pi \mu^2_K}{3}\int_0^1
dxdy\int_0^{\infty}bdb hdh \alpha_{s}(\mu_f)\hat{T}(x,{\bf b};y,{\bf
h};\mu_f) S_t(x_i)S_t(y_j)\times\nonumber\\
& &\left[y \hat{\Psi}_{p}(x,b;\mu_f) \hat{\Psi}^*_{p}
(y,h;\mu_f)+(1+\bar{y})
\hat{\Psi}_{p}(x,b;\mu_f)\frac{\hat{\Psi}^{\prime *}_{\sigma}
(y,h;\mu_f)}{6}+\right.\nonumber\\
& &\left. \hat{\Psi}_{p}(x,b;\mu_f)\frac{\hat{\Psi}^*_{\sigma}
(y,h;\mu_f)}{2} \right] \exp\left[-S(x_i,y_j,Q,\mathbf{b},
\mathbf{h};\mu_f)\right] \ ,
\end{eqnarray}
where $\bar{x}=(1-x)$, $\bar{y}=(1-y)$ and $\hat{\psi}^{\prime
*}_{\sigma} (y,h;\mu_f)=\partial \hat{\psi}^{*}_{\sigma}
(y,h;\mu_f)/\partial y$. The hard scattering amplitude
$\hat{T}(x,{\bf b};y,{\bf h};\mu_f)$ is given by
\begin{eqnarray}
\hat{T}(x,{\bf b};y,{\bf h};\mu_f)&=&
K_0\left(\sqrt{\bar{x}\bar{y}}Qb\right)\Big(\theta(b-h)K_0
\left(\sqrt{\bar{y}}Qb\right)I_0
\left(\sqrt{\bar{y}}Qh\right)+\nonumber\\
& & \theta(h-b)K_0 \left(\sqrt{\bar{y}}Qh\right)I_0
\left(\sqrt{\bar{y}}Qb\right) \Big),
\end{eqnarray}
where the higher power suppressed terms such as
$(\mathbf{k_\perp}^2/Q^2)$ has been neglected in the numerator,
$I_0$ and $K_0$ are the modified Bessel functions of the first kind
and the second kind respectively. To ensure that the pQCD approach
is really applicable, one has to specify carefully the
renormalization scale $\mu_f$ in the strong coupling constant. Here
we take the scheme that is proposed in Refs.\cite{lis}, i.e. its
value is taken as the largest renormalization scale associated with
the exchanged virtual gluon in the longitudinal and transverse
degrees,
\begin{equation}\label{scale}
\mu_f=\max(\sqrt{\bar{x}\bar{y}}Q,1/b,1/h).
\end{equation}

The full form of the kaon LC wave function have four helicity
components (Table. \ref{tab1}): namely,
\begin{equation}
\Psi=(\Psi_{\uparrow\uparrow},\Psi_{\uparrow\downarrow},
\Psi_{\downarrow\uparrow},\Psi_{\downarrow\downarrow}),\;\;\;\;
(\Psi=\Psi_p,\;\Psi_\sigma)
\end{equation}
By including the higher helicity components, Eq.(\ref{finalQ}) can
be improved as
\begin{eqnarray}\label{finalQh}
F^{twist3}_{K}(Q^2)&=&\frac{128\pi\mu_K^2 }{3}\int_0^1
dxdy\int_0^{\infty}bdb hdh \alpha_{s}(\mu_f)\times
\hat{T}(x,{\bf b};y,{\bf h};\mu_f) S_t(x_i)S_t(y_j)\times\nonumber\\
& &\left[y \sum_{\lambda_1\lambda_2}{\cal
P}(\hat\Psi_{p},\lambda_1,\lambda_2)+
\frac{(1+\bar{y})}{6}\sum_{\lambda_1\lambda_2}{\cal
P}(\hat\Psi'_{\sigma},\lambda_1,\lambda_2)+
\frac{1}{2}\sum_{\lambda_1\lambda_2}{\cal
P}(\hat\Psi_{\sigma},\lambda_1,\lambda_2) \right]\nonumber\\
&&\times\exp\left[-S(x_i,y_j,Q,\mathbf{b},\mathbf{h};\mu_f)\right],
\end{eqnarray}
where $\hat\Psi'_{\sigma}=\partial \hat\Psi_{\sigma}/\partial x$ and
$$\sum_{\lambda_1\lambda_2}{\cal
P}(\hat\Psi_{p},\lambda_1,\lambda_2)=(\hat\Psi^{*}_{p\uparrow\downarrow}
\hat\Psi_{p\uparrow\downarrow} +\hat\Psi^{*}_{p\downarrow\uparrow}
\hat\Psi_{p\downarrow\uparrow})-(\hat\Psi^{*}_{p\uparrow\uparrow}
\hat\Psi_{p\uparrow\uparrow}+ \hat\Psi^{*}_{p\downarrow\downarrow}
\hat\Psi_{p\downarrow\downarrow}),$$
$$\sum_{\lambda_1\lambda_2}{\cal
P}(\hat\Psi'_{\sigma},\lambda_1,\lambda_2)=(\hat\Psi^{*}_{p\uparrow\downarrow}
\hat\Psi'_{\sigma\uparrow\downarrow}
+\hat\Psi^{*}_{p\downarrow\uparrow}
\hat\Psi'_{\sigma\downarrow\uparrow})-(\hat\Psi^{*}_{p\uparrow\uparrow}
\hat\Psi'_{\sigma\uparrow\uparrow}+
\hat\Psi^{*}_{p\downarrow\downarrow}
\hat\Psi'_{\sigma\downarrow\downarrow}),$$
$$\sum_{\lambda_1\lambda_2}{\cal
P}(\hat\Psi_{\sigma},\lambda_1,\lambda_2)=(\hat\Psi^{*}_{p\uparrow\downarrow}
\hat\Psi_{\sigma\uparrow\downarrow}
+\hat\Psi^{*}_{p\downarrow\uparrow}
\hat\Psi_{\sigma\downarrow\uparrow})-(\hat\Psi^{*}_{p\uparrow\uparrow}
\hat\Psi_{\sigma\uparrow\uparrow}+
\hat\Psi^{*}_{p\downarrow\downarrow}
\hat\Psi_{\sigma\downarrow\downarrow}).$$ For the hard scattering
amplitude $\hat{T}(x,{\bf b};y,{\bf h};\mu_f)$, we have implicitly
adopted the approximate relation, i.e. $\hat{T}(x,{\bf b};y,{\bf
h};\mu_f)^{\uparrow\uparrow+\downarrow\downarrow} \approx
-\hat{T}(x,{\bf b};y,{\bf
h};\mu_f)^{\uparrow\downarrow+\downarrow\uparrow} $, since it can be
found that
\begin{equation}\label{apphel}
\hat{T}(x,{\bf b};y,{\bf h};\mu_f)^{\uparrow\uparrow
+\downarrow\downarrow}=-\hat{T}(x,{\bf b};y,{\bf
h};\mu_f)^{\uparrow\downarrow+\downarrow\uparrow}+{\cal O}(1/Q^2).
\end{equation}

\section{Numerical results for the kaon electromagnetic form factor}

Based on the formulae presented in the last section, we discuss
sequentially the leading and the power suppressed contributions to
the kaon electromagnetic form factor within the space-like region in
the following. The differences between the pion and kaon
electromagnetic form factors shall also be discussed in the due
places. In the numerical calculations, we use
$\Lambda^{(n_f=4)}_{\over{MS}}=250MeV$. As for the phenomenological
parameter $\mu_K$, which is a scale characterized by the chiral
perturbation theory, we take its value to be $\mu_K \simeq 1.70$
GeV. And for definiteness, we take the conventional values for the
constitute quark masses: $m_{u,d}=0.30 {\rm GeV}$ and $m_s=0.45 {\rm
GeV}$.

\subsection{LC wave function of the kaon}

In order to obtain the numerical results for the kaon
electromagnetic form factor, we need to know its LC wave functions.
One useful way is to use the approximate bound state solution of a
hadron in terms of the quark model as the starting point for
modeling the hadronic valence wave function. In combination of the
spin-space wave function $\chi$ that comes from the Wigner rotation
\cite{wigner}, the full form of the kaon LC wave function can be
written as, $\Psi(x,\mathbf{k_\perp,\lambda})
=\varphi(x,\mathbf{k_\perp}) \chi$. The explicit form of the
spin-space wave function $\chi$ can be found in TAB.\ref{tab1}. As
for the radial part of the wave function, we adopt the model
constructed in Refs.\cite{wuhuang1,wuhuang2}, which is based on the
BHL-prescription \cite{bhl},
\begin{eqnarray}
\varphi_K(x,\mathbf{k}_{\perp}) &=& [1+B_K C^{3/2}_1(2x-1)+C_K
C^{3/2}_2(2x-1)]\frac{A_K}{x(1-x)}\nonumber\\
&& \times\exp \left[-\beta_K^2 \left(\frac{k_\perp^2+m_1^2}{x}+
\frac{k_\perp^2+m_2^2} {1-x}\right)\right],\label{waveK} \\
\varphi_{\sigma}(x,\mathbf{k}_\perp) &=& A_\sigma \exp
\left[-\beta_K^2 \left(\frac{k_\perp^2+m_1^2}{x}+
\frac{k_\perp^2+m_2^2} {1-x}\right)\right],\label{waveS}\\
\varphi_p(x,\vec{k_\perp}) &=& [1+B_p
C^{1/2}_1(2x-1)+C_p C^{1/2}_2(2x-1)]\frac{A_p}{x(1-x)}\nonumber\\
&&\times \exp \left[-\beta_K^2 \left(\frac{k_\perp^2+m_1^2}{x}+
\frac{k_\perp^2+m_2^2} {1-x}\right)\right],\label{waveP}
\end{eqnarray}
where $m_1=m_u$ and $m_2=m_s$ for $F_{K^+}(Q^2)$, $m_1=m_d$ and
$m_2=m_s$ for $F_{K^0}(Q^2)$. $C^{3/2}_n(2x-1)$ and
$C^{1/2}_n(2x-1)$ are Gegenbauer polynomials. A more complicated
model that is also based on BHL-prescription is suggested in
Ref.\cite{maboqiang}. Numerically it can be found that the two model
wave functions behave very likely under the same constraints.
Additionally, it has argued in Ref.\cite{choi} that an extra factor
$\sqrt{\partial k_z/\partial x}$ should be included into the LC wave
function, otherwise, one can not obtain the right asymptotic
behavior for the pion electromagnetic form factor at $Q^2\to\infty$.
However, we have checked that without this factor, one can still
obtain the right power behavior for the pion electromagnetic form
factor as shown in Ref.\cite{piem2}, and such a factor will not
bring any new features for the LC wave function if its parameters
are determined properly. So we shall adopt the simpler form as
suggested in Refs.\cite{wuhuang1,wuhuang2} to do our calculation.

The four parameters $A_K$, $B_K$, $C_K$ and $\beta_K$ of
$\varphi_K(x,\mathbf{k}_{\perp})$ can be determined by the first two
Gegenbauer moments $a^K_1$ and $a^K_2$ of $\phi_K(x)$, the
constraint $\langle \mathbf{k}_\perp^2 \rangle^{1/2}_K \approx
0.350{\rm GeV}$ \cite{gh} and the normalization condition
\begin{eqnarray}
&&\int^1_0 dx \int_{k_\perp^2<\mu_0^2} \frac{d^{2}{\bf
k}_{\perp}}{16\pi^3}\frac{a_1 a_2-\mathbf{k}_\perp^2}
{[(a^2_1+\mathbf{k}^2_\perp) (a^2_2+\mathbf{k}^2_\perp)]^{1/2}}
\varphi_{K}(x,\mathbf{k_\perp}) \nonumber\\
&&=\int^1_0 dx \int_{k_\perp^2<\mu_0^2} \frac{d^{2}{\bf
k}_{\perp}}{16\pi^3}\frac{m_{1}(1-x)+m_{2}x}
{\sqrt{k^2_\perp+\left[m_{1}(1-x)+m_{2}x\right]^2}}
\varphi_{K}(x,\mathbf{k_\perp}) =\frac{f_K}{2\sqrt{6}},
\label{normalization}
\end{eqnarray}
where $\mu_0$ stands for some hadronic scale that is of order ${\cal
O}(1GeV)$. Here, the wave function is normalized to $f_K/2\sqrt{6}$
only for convenience, which is different from that of
Refs.\cite{wuhuang1,wuhuang2} that is normalized to one \footnote{It
is noted that the unit of $A_K$ in Ref.\cite{wuhuang1} should be
corrected from $GeV^{-1}$ to $GeV^{-2}$.}, where the factor
$f_K/2\sqrt{6}$ has been absorbed into the hard part of the $B\to K$
transition form factor. And the average value of the transverse
momentum square of kaon is defined as
\begin{displaymath}
\langle \mathbf{k}_\perp^2 \rangle_K=\frac{\int dx
d^2\mathbf{k}_\perp |\mathbf{k}_\perp^2| |\Psi_K(x,{\bf
k}_{\perp})|^2} {\int dx d^2\mathbf{k}_\perp |\Psi_K(x,{\bf
k}_{\perp})|^2}=\frac{\int dx d^2\mathbf{k}_\perp
|\mathbf{k}_\perp^2| |\varphi_K(x,{\bf k}_{\perp})|^2} {\int dx
d^2\mathbf{k}_\perp |\varphi_K(x,{\bf k}_{\perp})|^2} .
\end{displaymath}
The first Gegenbauer moment $a_1^K$ has been studied by the
light-front quark model \cite{quark1}, the LCSR approach
\cite{lcsr1,pballa1k,ballmoments,lenz,zwicky} and the lattice
calculation \cite{lattice1,lattice2,lattice3} and etc. The higher
Gegenbauer moments, such as $a^K_2$, are still determined with large
uncertainty \cite{sumrule,pballa1k,ballmoments,lcsr1,latt,instat}.
In the following calculation, if not specially stated, we take
$a^K_1(1{\rm GeV})=0.05$ \cite{lcsr1} and $a^K_2(1GeV)=0.115$
\cite{sumrule} to be their default values, and shall discuss the
uncertainties caused by these two factors in due places. By taking
these default values, we obtain
\begin{equation}
A_{K}=12.55GeV^{-1},\; B_{K}=0.0605,\; C_{K}=0.0348,\;
\beta_{K}=0.8706 GeV^{-1}\ .
\end{equation}

The parameter $A_\sigma$ of $\varphi_\sigma(x,\mathbf{k}_\perp)$ can
be determined by its normalization condition similar to
Eq.(\ref{normalization}), i.e. $A_\sigma=65.04 GeV^{-1}$ . And the
coefficients $A_p$, $B_p$ and $C_p$ of
$\varphi_p(x,\mathbf{k}_\perp)$ can be determined by the DA moments
of $\varphi_p(x,\mathbf{k}_\perp)$. To discuss the uncertainty
caused by $\Psi_p$, we take two groups of DA moments that have been
obtained in Refs.\cite{ballmoments,instaton} to determine the
coefficients $A_p$, $B_p$ and $C_p$, where the moments in
Ref.\cite{ballmoments} are derived by using the QCD light-cone sum
rules and the moments in Ref.\cite{instaton} are derived based on
the effective chiral action from the instanton:
\begin{eqnarray}
{\rm Group \;\;1\;\;\;\;} [25] : && \langle
x^0\rangle^K_p=1,\;\;\langle x^1\rangle^K_p=0.06124,\;\; \langle
x^2\rangle^K_p=0.36757 , \\
{\rm Group \;\;2\;\;\;\;} [36] : && \langle
x^0\rangle^K_p=1,\;\;\langle x^1\rangle^K_p=0.00678,\;\; \langle
x^2\rangle^K_p=0.35162.
\end{eqnarray}
Here the moments are defined as $\langle x^i\rangle^K_p =\int^1_{0}
dx (2x-1)^i\phi_p(1-x,\mu_0)$ with $i=(0,1,2)$. It should be noted
that the moments defined in Ref.\cite{ballmoments,instaton} are for
$\phi_p(1-x,\mu_0)$ other than $\phi_p(x,\mu_0)$, since in these
references $x$ stands for the momentum fraction of $s$-quark in the
kaon ($\bar{K}$), while in the present paper $x$ stands for the
momentum fraction of the light quark $q$ in the kaon ($K$). Taking
the above two groups of DA moments, the parameters of
$\varphi_p(x,\vec{k_\perp})$ can be determined as,
\begin{eqnarray}\label{group1}
{\rm Group \;\;1:\;\;}&&\;\;\;A^1_p=12.12 {\rm GeV}^{-1},\;\;\;
B^1_p=0.3062,\;\;\; C^1_p=1.604 ,\\
\label{group2} {\rm Group \;\;2:\;\;}&&\;\;\;A^2_p=12.04 {\rm
GeV}^{-1},\;\;\; B^2_p=0.4711,\;\;\; C^2_p=1.506 .
\end{eqnarray}
It is found that both distribution amplitudes are double humped
curves and are highly suppressed in the endpoint region. Such
feature is necessary to suppress the endpoint singularity coming
from the hard-scattering kernel and then to derive a more reasonable
results for the twist-3 contributions to the kaon electromagnetic
form factor.

\subsection{Valence Fock state contribution in the low energy region}

At the present, the experimental data on the kaon electromagnetic
form factor are concentrated in the low energy region $Q^2<1GeV^2$,
c.f. Ref.\cite{pdg}. The soft part contribution can be written as
\begin{equation}\label{gensoft}
F^s_{K}(Q^2)=\sum_{\lambda,\lambda^{\prime}}\sum_{j}e_j\int^1_0 dx
\int \frac{d^{2}{\bf k}_{\perp}}{16\pi^3}\Psi_{K}^{*}(x,{\bf
k_{\bot}},\lambda)\Psi_{K} (x,{\bf k'_{\bot}},\lambda^{\prime})
+\cdots,
\end{equation}
where $\lambda$, $\lambda^{\prime}$ are the helicities of the wave
function respectively, and the first term is the lowest order
contribution from the minimal Fock state (valence Fock state) and
the ellipses represent those from higher Fock states, which are down
by powers of $1/Q^2$ and by powers of $\alpha_s$ in the large $Q^2$
region. In general, the kaon electromagnetic form factor should sum
over all of higher Fock state contributions in the low energy
region. If only taking the leading-twist LC wave function of the
valence Fock state, we can examine the contribution from the valence
Fock state in the low energy region, i.e.
\begin{equation}
\label{softkc} F^{s(V)}_{K}(Q^2)=e_1\left[\int^1_0 dx \int
\frac{d^{2}{\bf k}_{\perp}}{16\pi^3}\kappa
\varphi_K(x,\mathbf{k}_\perp) \varphi_K^{*}(x,\mathbf{k'}_\perp)
\right]+ e_2\Bigg[m_1\leftrightarrow m_2\Bigg],
\end{equation}
where $e_1=2/3$, $e_2=1/3$, $m_1=m_u$ and $m_2=m_s$ for
$F^{s(V)}_{K^+}$ and $e_1=-1/3$, $e_2=1/3$, $m_1=m_d$ and $m_2=m_s$
for $F^{s(V)}_{K^0}$ respectively. The terms in the second bracket
is obtained by transforming the terms in the first bracket through
$m_1\leftrightarrow m_2$, the coefficient $\kappa$ that is from the
spin-space Winger rotation can be written as
\begin{eqnarray}
\kappa &=& \frac{(a_1 a_2-\mathbf{k}_\perp^2) (a'_1
a'_2-\mathbf{k'}_\perp^{2})+(a_1+a_2) (a'_1 + a'_2)\mathbf{k}_\perp
\cdot \mathbf{k'}_\perp} {[(a^2_1+\mathbf{k}^2_\perp)
(a^2_2+\mathbf{k}^2_\perp)(a^{\prime 2}_1+
\mathbf{k'}^2_\perp)(a^{\prime
2}_2+\mathbf{k'}^2_\perp)]^{1/2}}\nonumber \\
&=&\frac{\left[m_{1}(1-x)+m_{2}x\right]^2+ \mathbf{k}_\perp \cdot
\mathbf{k'}_\perp} {\sqrt{k^2_\perp +\left[m_{1}(1-x)
+m_{2}x\right]^2} \sqrt{k^{\prime 2} _\perp+\left[m_{1}(1-x)
+m_{2}x\right]^2}},
\end{eqnarray}
where ${\bf k'}_{\perp}={\bf k}_{\perp}+(1-x){\bf q}_{\perp}$ for
the final state LC wave function when taking the Drell-Yan-West
assignment \cite{dy}. Since $m_1\neq m_2$, we have
$F^{s(V)}_{K^0}(Q^2)\neq 0$, which is different from the pionic
case, i.e. $F^{s(V)}_{\pi^0}(Q^2)\equiv0$ because $\pi^0$ has equal
quark masses.

We proceed to integrate the transverse momentum $\mathbf{k}_{\bot}$
in Eq.(\ref{softkc}) with the help of the Schwinger
$\alpha-$representation method,
\begin{equation}\label{schwinger}
\frac{1}{A^{\kappa}}=\frac{1}{\Gamma(\kappa)}\int_0^{\infty}
\alpha^{\kappa-1}e^{-\alpha A}d\alpha\ .
\end{equation}
Doing the integration over $\mathbf{k}_{\perp}$, we obtain
\begin{eqnarray}
&&\int^1_0 dx \int \frac{d^{2}{\bf k}_{\perp}}{16\pi^3}\kappa
\varphi_K(x,\mathbf{k}_\perp) \varphi_K^{*}(x,\mathbf{k'}_\perp)\nonumber\\
&=&\int^1_0 dx\int^\infty_0 d\lambda \frac{A_K^2(x)}{128\pi^2 ( 1 +
\lambda)^3}\times\left[ I_0\left(\frac{Q^2( x-1)\beta_K^2 \lambda^2}
{4x( 1 + \lambda)}\right)\bigg( \frac{4(1-x)x( 1 +
\lambda)}{\beta_K^2} \right.\nonumber\\
& &  \left. -Q^2( 1 - x )^2 ( 2 + \lambda ( 4 +\lambda )) + 8m_b^2(1
+\lambda)^2\bigg) - I_1\left(\frac{Q^2(x -1)\beta_K^2 {\lambda }^2}
{4x( 1 + \lambda) }\right)Q^2(1-x)^2 \lambda^2
\right]\nonumber\\
& &\times\exp\left[-\frac{\beta_K^2[8m_a^2(1 + \lambda)
+8m^2_b\lambda(\lambda+1)+ Q^2(1-x)^2( 2 + \lambda( 4 +
\lambda))]}{4(1-x ) x ( 1 + \lambda) }\right], \label{probability}
\end{eqnarray}
where the short notations $A_K(x)\equiv\frac{A_K}{x(1-x)}[1+B_K
C^{3/2}_1(2x-1)+C_K C^{3/2}_2(2x-1)]$, $m_a^2=m_1^{2} (1-x)+m_2^{2}
x$, $m_b^2=\left[m_{1}(1-x)+m_{2}x\right]^2$, and $I_n\,\,(n=0,1)$
is the modified Bessel function of the first kind. Substituting
Eq.(\ref{probability}) into Eq.(\ref{softkc}) and doing the
expansion in the small $Q^2$ limit, we obtain
\begin{eqnarray}
&&F^{s(V)}_{K}(Q^2)|_{Q^2=0}\nonumber\\
&=&e_1\left\{\int^1_0 dx\int^\infty_0 d\lambda \frac{A_K^2(x)}
{16\pi^2 ( 1 + \lambda)^2}\exp\left[-\frac{2\beta_K^2 ( m_a^2 +
m^2_b\lambda)}{(1-x)x }\right]\left[m^2_b(1
+\lambda)+\frac{x(1-x)}{2\beta_K^2}\right]\right\}\nonumber\\
&& +e_2\Bigg\{m_1\leftrightarrow m_2\Bigg\} ,
\end{eqnarray}
where the term $[m^2_b(1 +\lambda)]$ in the second square bracket
comes from the ordinal helicity components, while the remaining
terms are from the higher helicity components. As for the mean
square radius $\langle r^2_{K}\rangle_V$, we obtain
\begin{eqnarray}\label{radius}
\langle r_{K}^2 \rangle_V &\approx&-6\left.\frac{\partial
F^{s(V)}_{K}(Q^2)}{\partial Q^2}\right|_{Q^2=0}\nonumber\\
&=&e_1\left\{\int^1_0 dx\int^\infty_0 d\lambda\frac{3
A_K^2(x)\beta_K^2}{32 {\pi }^2 x ( 1 +
\lambda)^3}\exp\left[-\frac{2\beta^2_K(m_a^2 +
\lambda m_b^2)}{(1-x) x}\right]( 1-x )( 2 + 4\lambda+\lambda^2 )\right.\nonumber\\
& &\left.\times\Bigg[ \frac{(1-x) x}{\beta_K^2}+m_b^2( 1 +
\lambda)\Bigg]\right\}+ e_2\Bigg\{m_1\leftrightarrow m_2\Bigg\} .
\end{eqnarray}

\begin{figure}
\centering
\includegraphics[width=0.50\textwidth]{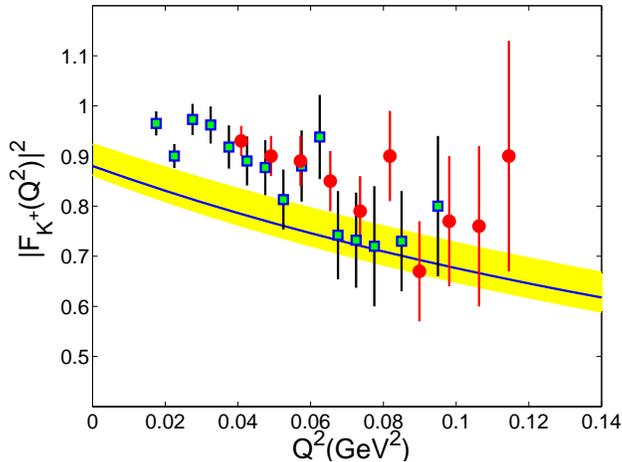}
\caption{The valence Fock state contribution to the kaon
electromagnetic form factor $|F_{K^+}(Q^2)|^2$ in the low energy
region, where the experimental data is taken from Ref.\cite{cjb}.
The shaded band is obtained with $a^K_1(1GeV)\in[0.03, 0.07]$ and
$a^K_2(1GeV)\in[0.05, 0.10]$, and the solid line is for
$a^K_1(1GeV)=0.05$ and $a^K_2(1GeV)=0.115$. } \label{form}
\end{figure}

The result for the soft contribution to the kaon electromagnetic
form factor is shown in Fig.(\ref{form}), where the solid line is
for $a^K_1(1GeV)=0.05$ and $a^K_2(1GeV)=0.115$ and the shaded band
is obtained with $a^K_1(1GeV)\in[0.03, 0.07]$ and
$a^K_2(1GeV)\in[0.05, 0.10]$. The valence quark contribution is
slightly below the experimental data, which means that there are
still some space for the higher Fock state contributions. With the
help of Eq.(\ref{probability}), we can estimate the probability of
finding the valance states in the charged/neutral kaon, e.g. we
obtain $(P_{u\bar{s}}=0.901<1.0)$, which shows that higher Fock
states and higher twist terms should also be considered to give a
full understanding of the form factor at the energy region
$Q^2\to0$. Such probability can be further divided into two parts:
$(P_{u\bar{s}}^{(\lambda_1+\lambda_2=0)}=0.562)$ for the usual
helicity components and $(P_{u\bar{s}}^{(\lambda_1+\lambda_2=\pm
1)}=0.339)$ for the higher helicity states. It shows that the higher
helicity components have the same importance as that of the usual
helicity components in the soft energy region. It is noted that the
higher helicity components' contribution to the kaon electromagnetic
form factor has also been studied with the LC framework in
Ref.\cite{maboqiang}, where the probability of the leading Fock
state is just normalized to one and the experimental data on the
mean-square radius of charged/neutral kaon are used to determine the
wave function parameters. As argued above, this simple treatment
maybe not right, since then the contribution from the valence state
can be enhanced and become important inadequately \footnote{As has
been pointed out in Ref.\cite{piem2}, the condition for the pionic
case is more serious, where the probabilities for the value quark
state is only about $74\%$.}.

As for the charged and neutral mean square radii $\langle
r^2_{K^+}\rangle_{V}$ and $\langle r^2_{K^0}\rangle_{V}$, we obtain
$\langle r^2_{K^{\pm}}\rangle_{V}^{1/2}=0.570\; {\rm fm}$ and
$\langle r^2_{K^0}\rangle_{V}=-0.0736\; {\rm fm}^2$, which is
consistent with the Ref.\cite{radii}, while experimentally we have
$\langle r^2_{K^{\pm}}\rangle^{1/2} =0.560\pm0.031 \;{\rm fm}$ and
$\langle r^2_{K^0}\rangle=-0.077\pm0.010 \;{\rm fm}^2$ \cite{pdg}.
Further more, we give a simple estimation of the uncertainties
caused by the two Gegenbauer moments of the kaon twist-2 wave
function, e.g. by taking $a^K_1(1GeV)=0.05\pm0.02$ and
$a^K_2(1GeV)=0.10\pm 0.05$, we obtain the probability of finding the
valence Fock state in the kaon
$P_{u\bar{s}}=0.901^{+0.026}_{-0.010}$, and the uncertainties of
radii $\langle r^2_{K^{\pm}}\rangle_{V}^{1/2}=
0.570^{+0.021}_{-0.028} \;{\rm fm}$ and $\langle
r^2_{K^0}\rangle_{V}=-0.0736^{+0.018}_{-0.014} \;{\rm fm}^2$. It
should be noted that by taking different values for $a^K_1$ and
$a^K_2$, all the undermined parameters of the wave function should
be varied accordingly.

\subsection{leading twist contribution to the kaon electromagnetic form factor}

\begin{figure}
\centering
\includegraphics[width=0.49\textwidth]{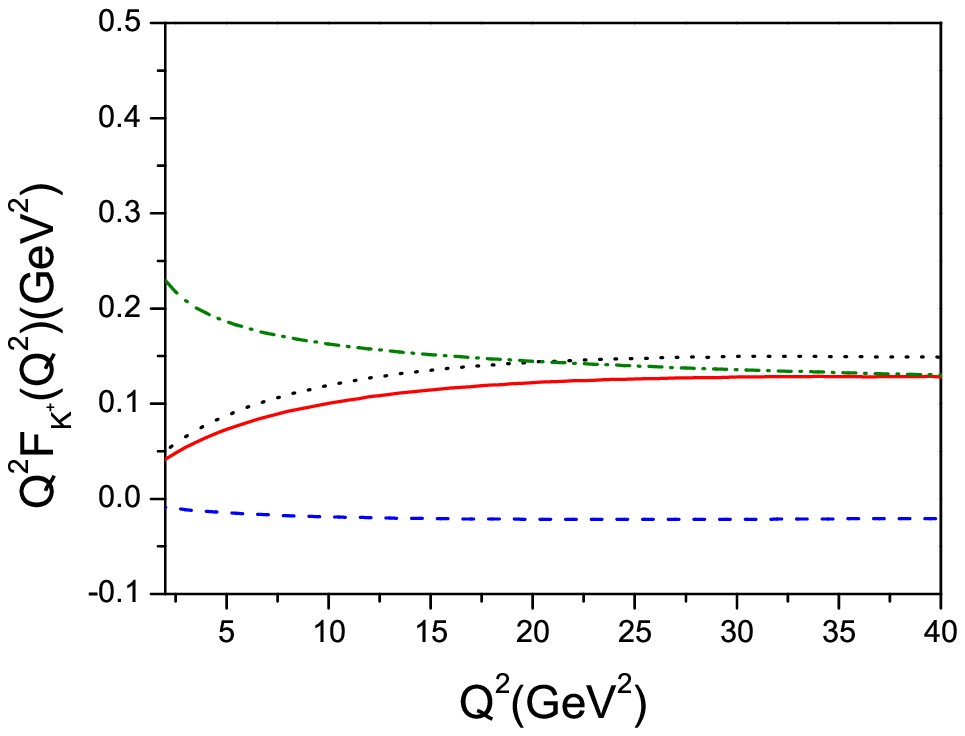}
\includegraphics[width=0.49\textwidth]{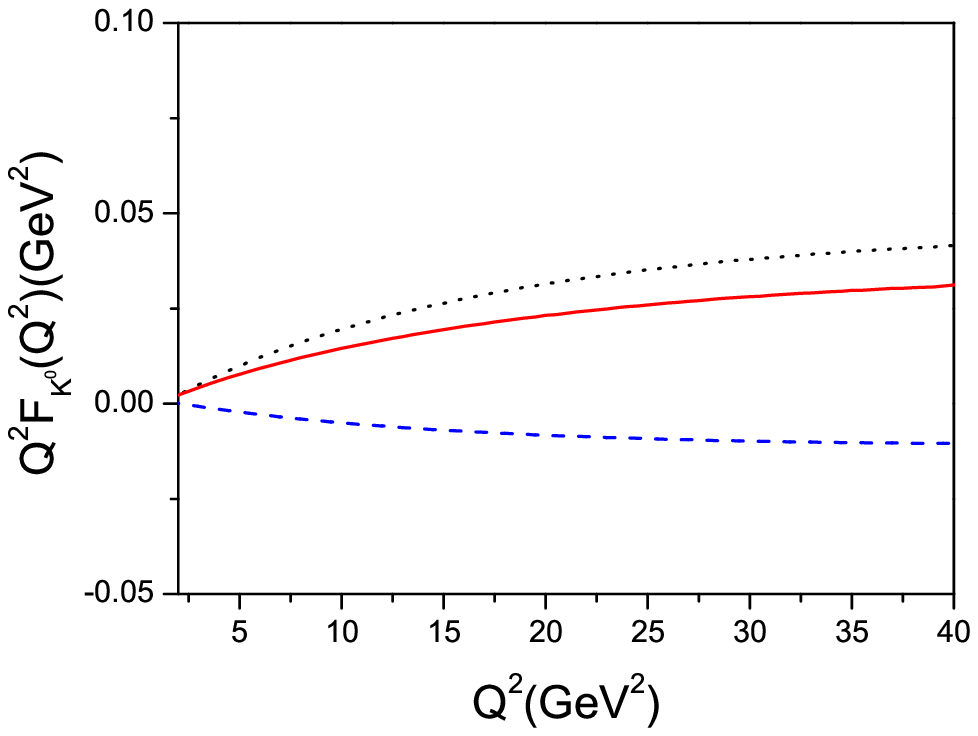}%
\caption{Leading twist contribution to the kaon electromagnetic form
factor in the intermediate and large energy region, where the left
is for $Q^2F_{K^+}(Q^2)$ and the right is for $Q^2F_{K^0}(Q^2)$. The
dotted line stands for the contribution from the usual helicity
($\lambda_1+\lambda_2=0$) components, the dashed line stands for the
contribution from the higher helicity ($\lambda_1+\lambda_2=\pm 1$)
components and the solid line is the total hard contribution, which
is the combined result for all the helicity components. The dash-dot
line stands for the usual asymptotic result of charged kaon.}
\label{hard}
\end{figure}

With the help of the LC wave function Eq.(\ref{waveK}), we show the
leading twist contribution to the kaon electromagnetic form factor
in the intermediate and large energy region in Fig.(\ref{hard}),
where the contribution from the usual helicity component or from the
higher helicity components are considered. It is shown that the
higher helicity components always suppress the usual helicity
components' contributions to the kaon electromagnetic form factor.
The usual asymptotic result of charged kaon, i.e.
$Q^2F_{K^+}(Q^2)|_{asy}=8\pi f_K^2\alpha_s(Q^2)$, is also presented
in Fig.(\ref{hard}) for reference. It can be found that the leading
contribution of the hard-scattering amplitude from the higher
helicity components is of order $1/Q^4$, which is next-to-leading
contribution compared to the contribution coming from the ordinary
helicity component, but it shall give sizable contributions to the
kaon electromagnetic form factor in the intermediate energy region.
The net contribution shows the right power behavior
$Q^2F_{K^+}(Q^2)|_{Q^{2}\to\infty}\to {\rm const}$. In the present
work, we have considered the $k_T$ dependence both in the wave
function and in the hard scattering amplitude consistently within
the LC pQCD approach, then our results present a right power
behavior for the higher helicity components' contributions.
Secondly, in contrary to the pionic case, we obtain
$F_{K^0}(Q^2)=-F_{\bar{K}^0}(Q^2)\neq 0$ at $Q^2\neq 0$, which are
rightly caused by the $SU_f(3)$-breaking effect and are strongly
dependent on the constitute quark masses.

\begin{figure}
\centering
\includegraphics[width=0.49\textwidth]{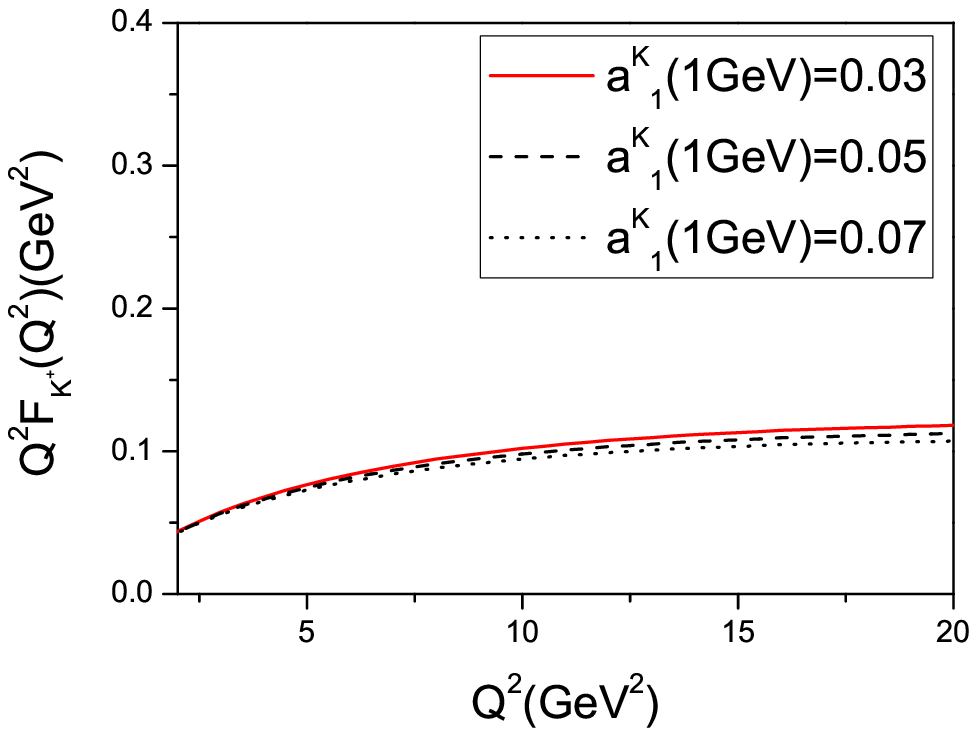}
\includegraphics[width=0.49\textwidth]{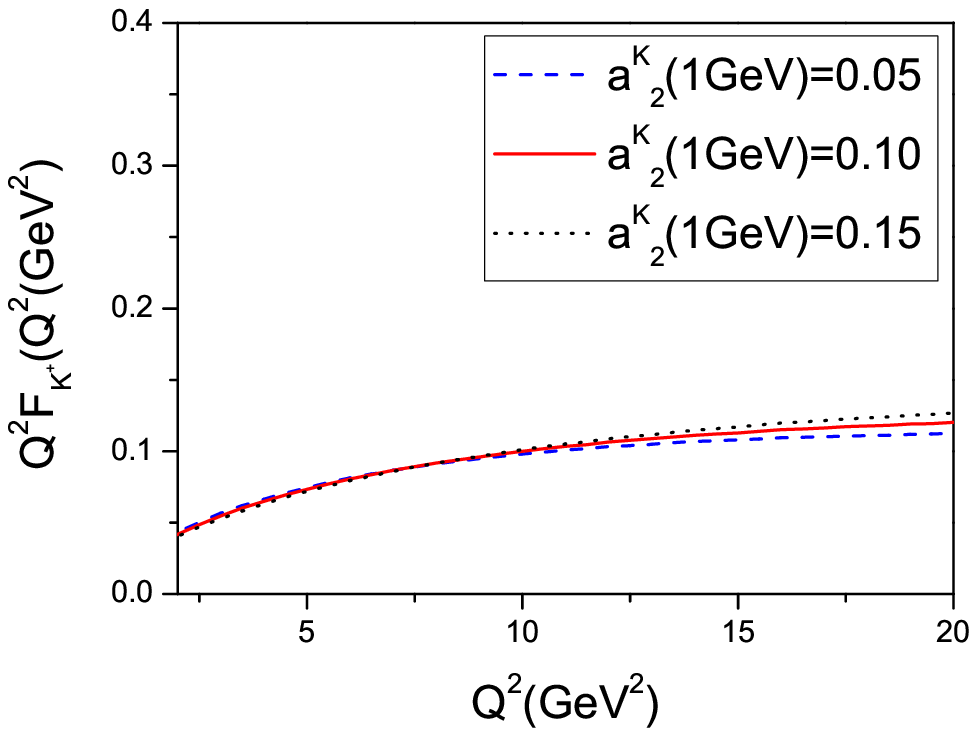}%
\caption{Uncertainties of the leading twist contribution to
$Q^2F_{K^+}(Q^2)$ caused by $a^K_1$ and $a^K_2$, where the left
diagram is for fixed $a^K_2(1GeV)=0.05$ with $a^K_1(1GeV)=0.03$,
$0.05$ and $0.07$, and the right diagram is for fixed
$a^K_1(1GeV)=0.05$ with $a^K_2(1GeV)=0.05$, $0.10$ and $0.15$
respectively.} \label{twist2un}
\end{figure}

We take the charged kaon electromagnetic form factor as a concrete
example to show the uncertain caused by $a^K_1(1GeV)$ and
$a^K_2(1GeV)$, which are varied within the region of $[0.03, 0.07]$
and $[0.05, 0.15]$ respectively. We draw the charged kaon
electromagnetic form factor in Fig.(\ref{twist2un}), where the left
diagram is for fixed $a^K_2(1GeV)=0.05$ with $a^K_1(1GeV)=0.03$,
$0.05$ and $0.07$, and the right diagram is for fixed
$a^K_1(1GeV)=0.05$ with $a^K_2(1GeV)=0.05$, $0.10$ and $0.15$
respectively. $Q^2F_{K^+}(Q^2)$ decreases with the increment of
$a^K_1$. From Fig(\ref{twist2un}), it can be found that the
uncertainty of the form factor caused by $a^K_1(1{\rm GeV})=
0.05\pm0.02$ is small, e.g. it is about $\pm 5\%$ for
$q^2\in[2,20]GeV^2$. And the uncertainty of the form factor caused
by $a^K_2(1{\rm GeV})$ varying within a bigger region $[0.05,0.15]$
is also small, i.e. which is about $4\%-9\%$ for
$q^2\in[2,20]GeV^2$. $Q^2F_{K^+}(Q^2)$ decreases with the increment
of $a^K_2$ in the lower energy region $q^2\lesssim 6GeV^2$ and
increases with the increment of $a^K_2$ in the higher energy region
$q^2 \gtrsim 6GeV^2$.

\subsection{twist-3 contribution to the kaon electromagnetic form factor}

\begin{figure}
\centering
\includegraphics[width=0.49\textwidth]{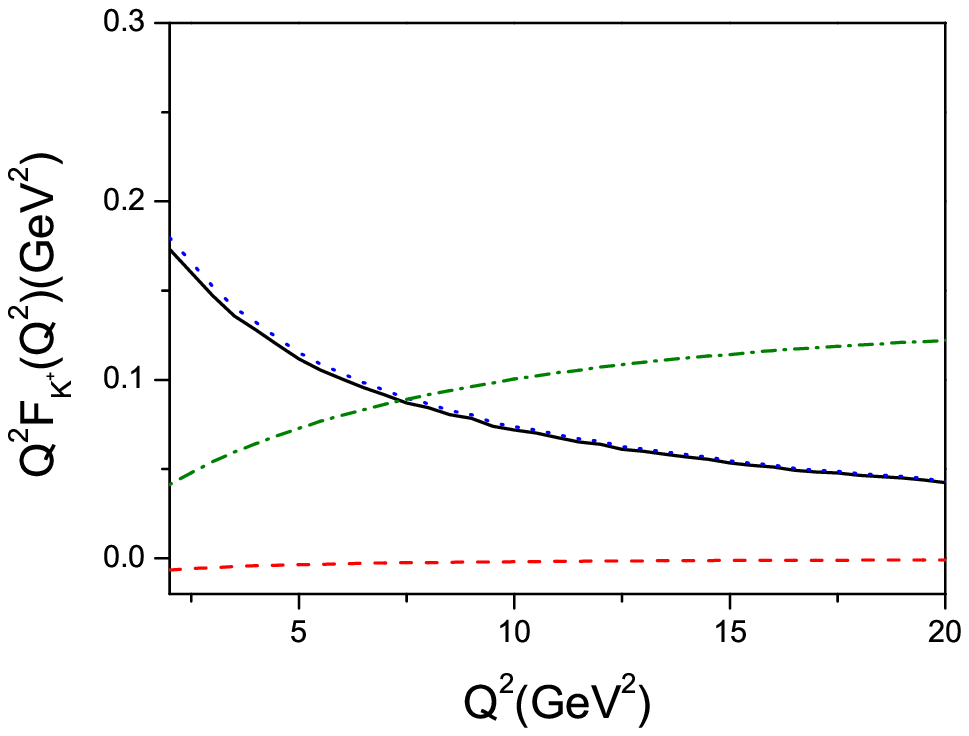}
\includegraphics[width=0.49\textwidth]{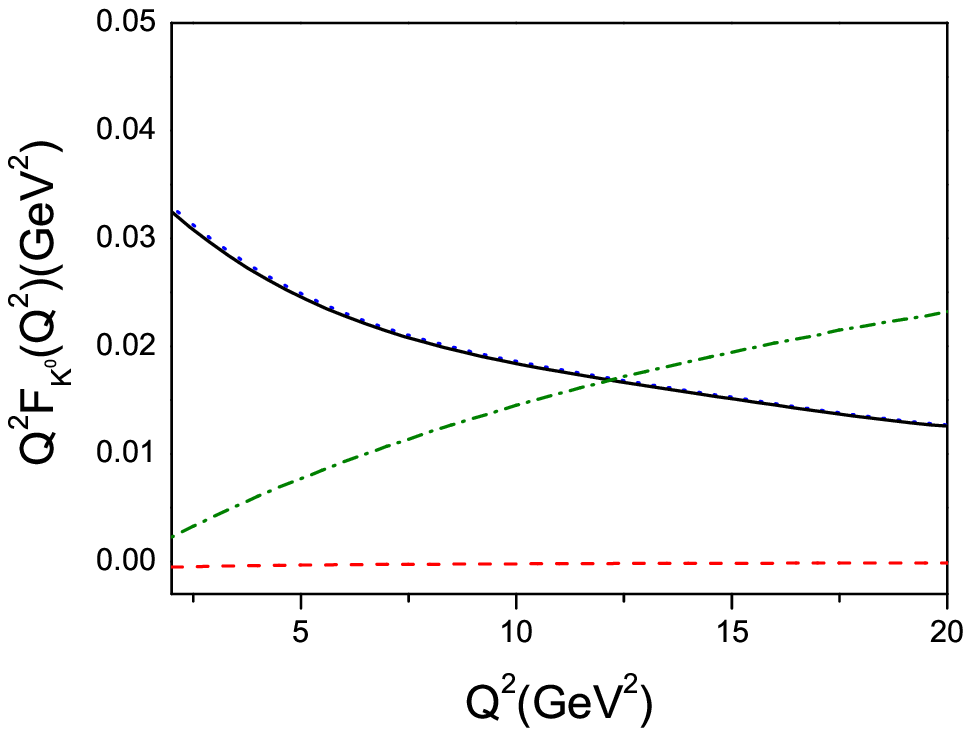}%
\caption{Twist-3 contribution to the kaon electromagnetic form
factor $Q^2F_{K^+}(Q^2)$ and $Q^2F_{K^0}(Q^2)$. The dotted line
stands for the contribution from the usual helicity
($\lambda_1+\lambda_2=0$) components, the dashed line stands for the
contribution from the higher helicity ($\lambda_1+\lambda_2=\pm 1$)
components and the solid line is the total hard contribution, which
is the combined result for all the helicity components. As a
comparison, the twist-2 contribution is shown in dash-dot line.}
\label{twista}
\end{figure}

We show the twist-3 contribution to kaon electromagnetic form
factors $Q^2F_{K^+}(Q^2)$ and $Q^2F_{K^0}(Q^2)$ in
Fig.(\ref{twista}), which are obtained with the full form of the LC
wave functions $\Psi^f_p(x,\mathbf{k_\perp})$ and
$\Psi^f_\sigma(x,\mathbf{k_\perp})$ and with the Group 1 parameters
for $\Psi_p(x,\mathbf{k}_\perp)$. It is found that at the twist-3
level, the higher helicity components' contributions to the form
factor are negative and small in comparison to that of the usual
helicity components. The twist-2 contribution is also presented in
Fig.(\ref{twista}) for comparison. At the twist-3 level, both the
charged and the neutral kaon electromagnetic form factors decrease
with the increment $Q^2$, and the charged form factor becomes
smaller than the twist-2 contribution at around $Q^2=7GeV^2$, which
is changed to be $Q^2\simeq 12GeV^2$ for the neutral case. This
implies that the twist-3 contributions are sizable in the
intermediate energy region and are rightly power suppressed to the
twist-2 contributions in the large energy region, which is similar
to the pionic case as shown in Ref.\cite{piem1}.

\begin{figure}
\centering
\includegraphics[width=0.48\textwidth]{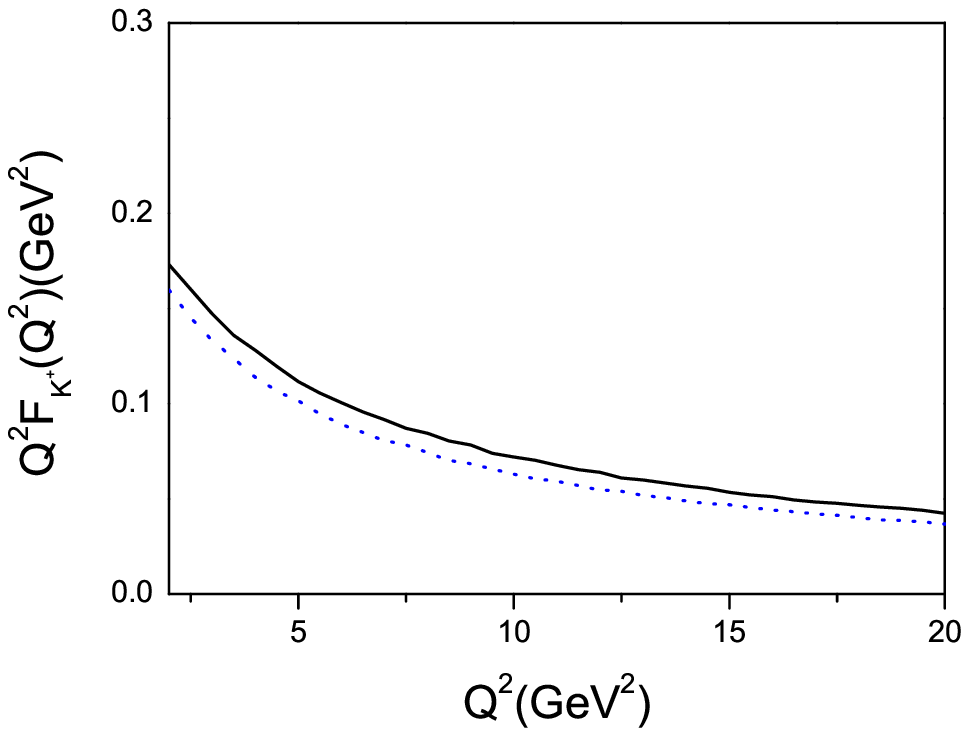}
\includegraphics[width=0.48\textwidth]{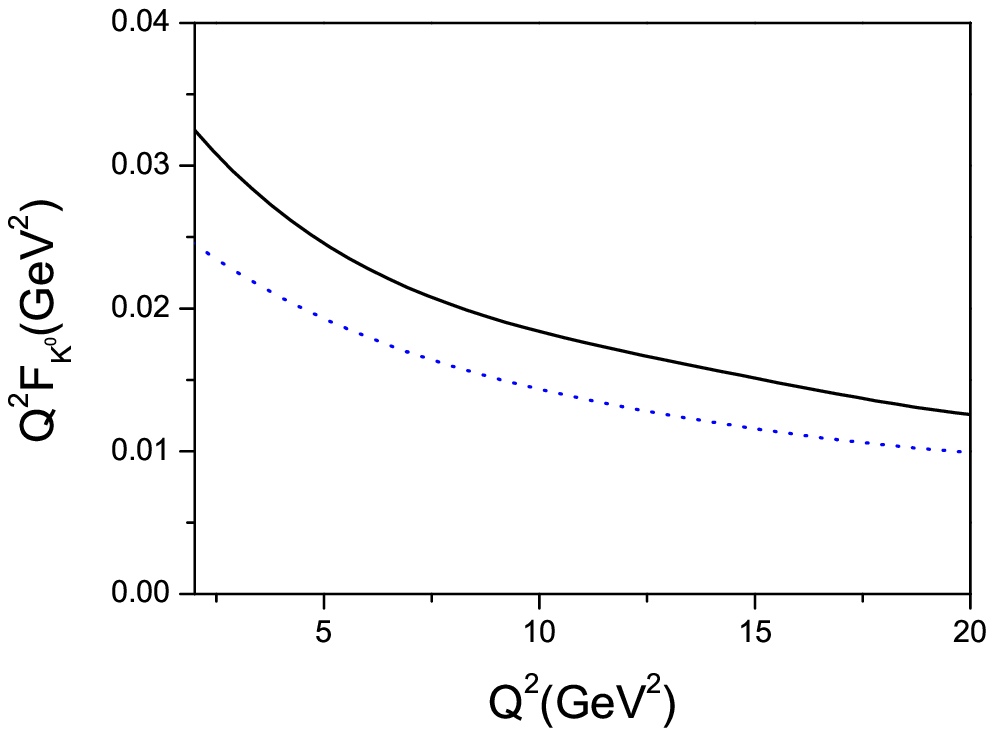}%
\caption{Uncertainty caused by two types of twist-3 wave function
$\Psi_p$ to the electromagnetic form factors $Q^2F_{K^+}(Q^2)$ (Left
diagram) and $Q^2F_{K^0}(Q^2)$ (Right diagram). The solid line and
the dotted line are for Group 1 and Group 2 parameters
respectively.} \label{twist3-1}
\end{figure}

\begin{figure}
\centering
\includegraphics[width=0.48\textwidth]{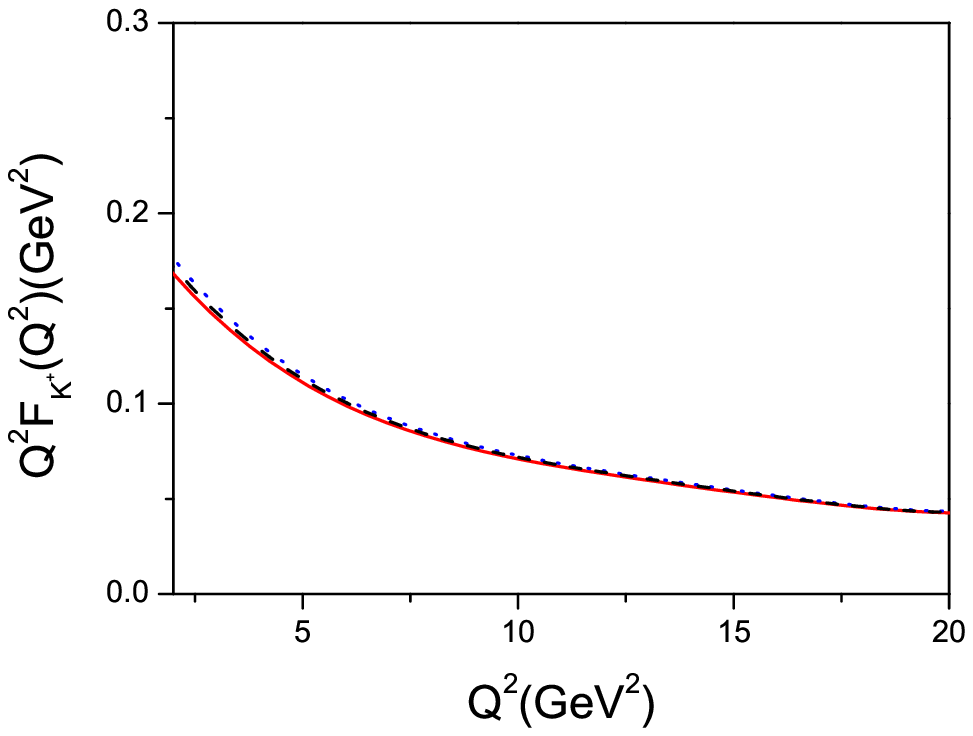}
\includegraphics[width=0.48\textwidth]{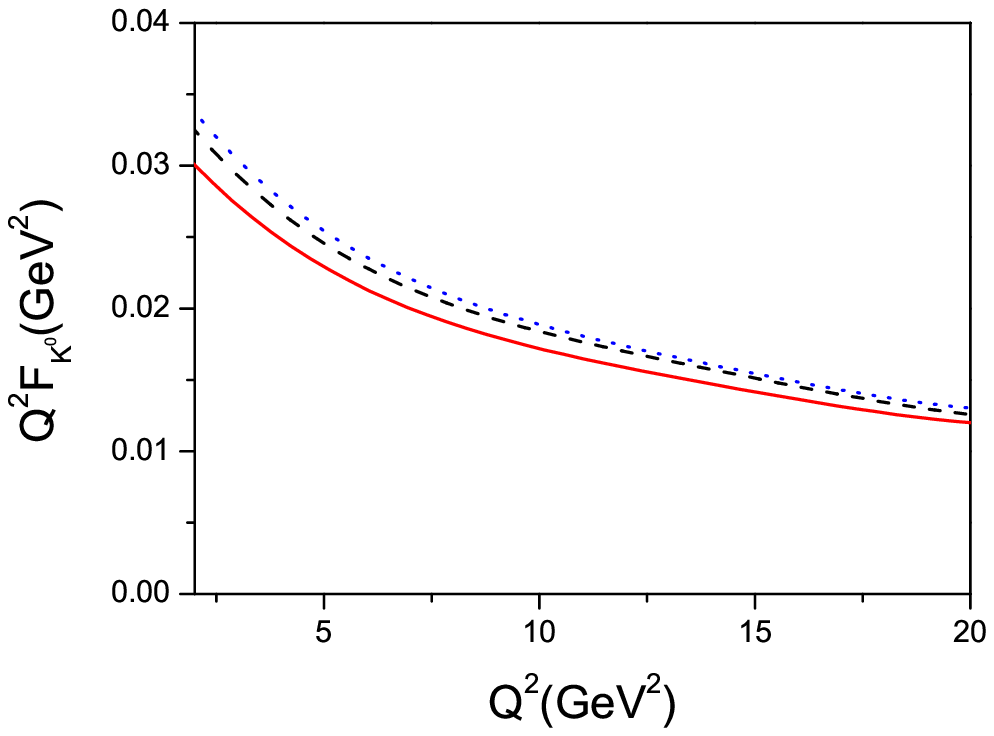}%
\caption{Uncertainty caused by the parameter $\beta_K$ of $\Psi_p$
with Group 1 parameters to the electromagnetic form factors
$Q^2F_{K^+}(Q^2)$ (Left diagram) and $Q^2F_{K^0}(Q^2)$ (Right
diagram). The solid line, the dashed line and the dotted line are
for $\beta_K=0.85GeV^{-1}$, $0.87GeV^{-1}$ and $0.89GeV^{-1}$
respectively.} \label{twist3-b}
\end{figure}

The main uncertainty sources for the twist-3 contribution come from
the wave function $\Psi_p(x,\mathbf{k}_\perp)$ and the parameter
$\beta_K$. We show the contributions to the charged kaon
electromagnetic form factor from the two groups of parameters for
$\Psi_p(x,\mathbf{k_\perp})$ in Fig.(\ref{twist3-1}), c.f.
Eqs.(\ref{group1}, \ref{group2}). It is found that the uncertainty
within the allowable energy region caused by these two groups of
parameters are about $10-20\%$ and $20-30\%$ for the charged case
and the neutral case respectively. Secondly, we show the uncertainty
caused by the parameter $\beta_K$ in Fig.(\ref{twist3-b}), where the
Group 1 moments \cite{ballmoments} are used to determine the
parameters of $\Psi(x,\mathbf{k}_\perp)$ and three typical values
$\beta_K=0.85GeV^{-1}$, $0.87GeV^{-1}$ and $0.89GeV^{-1}$ are
adopted \footnote{When varying $a^K_1(1GeV)$ and $a^K_2(1GeV)$
within the region of $[0.03, 0.07]$ and $[0.05, 0.15]$ respectively,
the value of $\beta_K$ shall vary within the region of
$[0.856,0.896]GeV^{-1}$.}. The twist-3 contribution increases with
the increment of $\beta_K$, and the uncertainty is less than $5\%$
for the charged form factor, while for the neutral form factor the
uncertainty changes to be $\sim 10\%$.

As for the higher order corrections, we present a naive estimation
of the next-to-leading order (NLO) twist-2 contribution to the
charged kaon electromagnetic form factor with the help of the
asymptotic DA, i.e. with the renormalization scale and the
factorization scale taken to be $\mu^2_R=\mu^2_f=Q^2$, it can
roughly be expressed as \cite{melic},
$Q^2F^{NLO}_{K^+}\approx(0.903GeV^2)\frac{f^2_K}{f^2_\pi}\alpha^2_s(Q^2)$.
Numerically the NLO correction will give about $\sim 20-30\%$ extra
contribution to the charged kaon electromagnetic form factor.

\section{Summary and conclusion}

\begin{figure}
\centering
\includegraphics[width=0.48\textwidth]{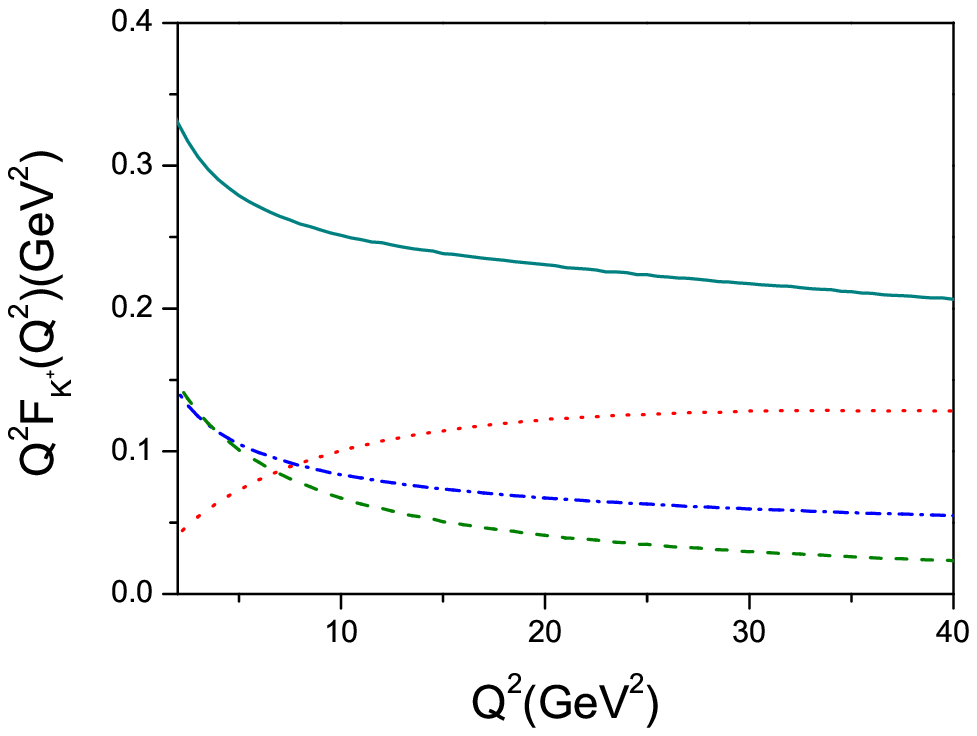}
\includegraphics[width=0.48\textwidth]{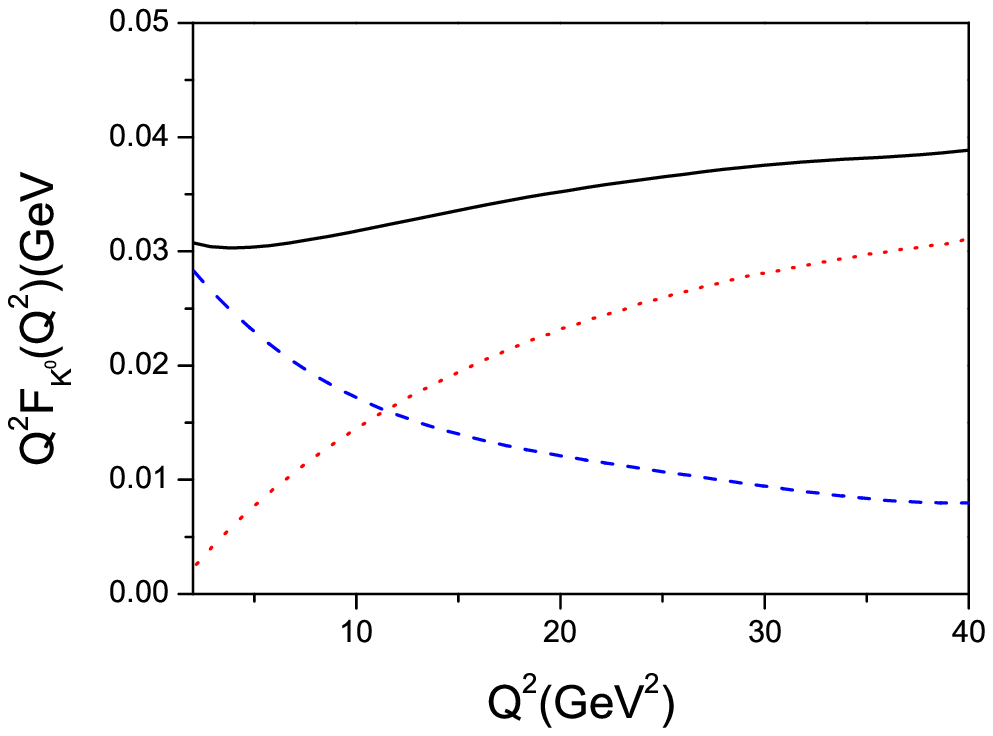}%
\caption{The combined hard contribution for the kaon electromagnetic
form factors $Q^2F_{K^+}(Q^2)$ (Left diagram) and $Q^2F_{K^0}(Q^2)$
(Right diagram). The solid line stands for the combined hard
contribution, the dotted line, the dashed line and the dash-dot line
are for LO twist-2, twist-3 and NLO twist-2 contributions
respectively.} \label{total}
\end{figure}

The $k_T$ factorization formalism provides a convenient framework
and has been widely applied to various processes. In this paper we
present a systematical study on the kaon electromagnetic form
factors $F_{K^{\pm},K^0,\bar{K}^0}(Q^2)$ within the $k_T$
factorization formalism. In order to get a deeper understanding of
the hard contributions at the energy region where pQCD is
applicable, we have examined the transverse momentum effects, the
contributions from the different helicity components and different
twist structures of the kaon LC wave function. Our results show that
the right power behavior of the hard contribution from the higher
helicity components and from the higher twist structures can be
obtained by keeping the $k_T$ dependence in the hard scattering
amplitude. The full estimation of the power suppressed contributions
to the kaon electromagnetic form factors $Q^2F_{K^+}(Q^2)$ and
$Q^2F_{K^0}(Q^2)$ is shown in Fig.(\ref{total}). The $k_T$
dependence in LC wave function affects the hard and soft
contributions substantially and the power-suppressed terms (twist-3
and higher helicity components) make an important contribution below
$Q^2\sim several\; GeV^2$ although they drop fast as $Q^2$
increasing.

The parameters of the proposed model wave function can be fixed by
the first two moments of its distribution amplitude and the
normalization condition. In this paper we have taken the first two
moments $a^K_1(1GeV)=0.05\pm0.02$ and $a^K_2(1GeV)=0.10\pm 0.05$. It
is found that the uncertainty of the kaon electromagnetic form
factor, which is caused by varying values within the above range, is
rather small. It is also found that the power-suppressed twist-3
contribution makes an important contribution at $Q^2\sim several\;
GeV^2$ and drops fast as $Q^2$ increasing. A naive estimation gives
the NLO correction about $\sim 20-30\%$ extra contribution to the
charged kaon form factor.

The relativistic effect due to the Wigner rotation have also been
applied to calculation the kaon electromagnetic form factor.
Consequently there are higher-helicity ($\lambda_1+\lambda_2=\pm 1$)
components in the spin space wave function besides the
usual-helicity ($\lambda_1+\lambda_2=0$) components. It is shown
that the higher helicity components have the same importance as that
of the usual helicity components for the soft energy region, e.g.
the probability of finding the valance states in the charged kaon
includes two parts: $(P_{u\bar{s}}^{(\lambda_1+\lambda_2=0)}=0.562)$
for the usual helicity components and
$(P_{u\bar{s}}^{(\lambda_1+\lambda_2=\pm 1)}=0.339)$ for the higher
helicity states for $a^K_1(1GeV)=0.05$ and $a^K_2(1GeV)=0.115$. By
taking $a^K_1(1GeV)=0.05\pm0.02$ and $a^K_2(1GeV)=0.10\pm 0.05$, we
obtain the uncertainty of the probabilities
$P_{u\bar{s}}=0.901^{+0.026}_{-0.010}$, It is found that the
hard-scattering amplitude for the higher-helicity components is of
order $1/Q^4$ which is the next to leading contribution compared
with the contribution coming from the ordinary helicity component,
but it can give sizable contributions to the kaon electromagnetic
form factor, especially for the twist-2 case as shown in
Fig.(\ref{hard}).

\begin{center}
\section*{Acknowledgements}
\end{center}

This work was supported in part by the Natural Science Foundation of
China (NSFC) and by the grant from the Chinese Academy of
Engineering Physics under the grant numbers: 2008T0401 and
2008T0402. \\

\end{document}